\documentclass[12pt,a4paper]{article}
\usepackage{amsfonts}
\advance\textheight by \topskip
\usepackage{cite}
\def\bmu{{\mathord{\hbox{\boldmath$\mu$}}}}
\def\bnu{{\mathord{\hbox{\boldmath$\nu$}}}}
\def\bomega{{\mathord{\hbox{\boldmath$\omega$}}}}
\def\bOmega{{\mathord{\hbox{\boldmath$\Omega$}}}}
\def\bphi{{\mathord{\hbox{\boldmath$\phi$}}}}
\def\bpsi{{\mathord{\hbox{\boldmath$\psi$}}}}
\def\brho{{\mathord{\hbox{\boldmath$\rho$}}}}
\def\bxi{{\mathord{\hbox{\boldmath$\xi$}}}}
\def\btheta{{\mathord{\hbox{\boldmath$\theta$}}}}

\def\i{\mathrm{i}}
\def\e{\mathrm{e}}
\def\today{13 July 1998}
\begin{document}
\thispagestyle{empty}
\def\thefootnote{\fnsymbol{footnote}}
\begin{flushright}
  hep-th/9807091\\
  DAMTP 1998-74 \\
  DTP-98/37
\end{flushright}
\vskip 2.5em
\begin{center}\LARGE
  A Local Logarithmic Conformal Field Theory
\end{center}\vskip 2em
\begin{center}\large
  Matthias R. Gaberdiel%
  $^{a}$\footnote{Email: {\tt M.R.Gaberdiel@damtp.cam.ac.uk}}
  and
  Horst G. Kausch%
  $^{b}$\footnote{Email: {\tt H.G.Kausch@dur.ac.uk}}%
\end{center}
\begin{center}\it$^a$
  Department of Applied Mathematics and Theoretical Physics, \\
  University of Cambridge, Silver Street, Cambridge CB3 9EW, U.K.
\end{center}
\begin{center}\it$^b$
  Department of Mathematical Sciences, \\
  University of Durham, South Road, Durham DH1 3LE, U.K.
\end{center}
\vskip 1em
\begin{center}
  \today
\end{center}
\vskip 1em
\begin{abstract}
  The local logarithmic conformal field theory corresponding to the
  triplet algebra at $c=-2$ is constructed. The constraints of
  locality and duality are explored in detail, and a 
  consistent set of amplitudes is found. The spectrum of the
  corresponding theory is determined, and it is found to be modular 
  invariant. This provides the first construction of a non-chiral
  rational logarithmic conformal field theory, establishing that such
  models can indeed define bona fide conformal field theories. 
  \begin{flushleft}
  \end{flushleft}
\end{abstract}

\setcounter{footnote}{0}
\def\thefootnote{\arabic{footnote}}

\section{Introduction}
\label{sec:intro}

Recently, a new class of chiral conformal field theories whose
correlation functions have logarithmic branch cuts has attracted some
attention. These models are believed to be important for the
description of certain statistical models, in particular in the theory
of (multi)critical polymers \cite{Sal92,Flohr95,Kau95,MNR97},
two-dimensional turbulence \cite{RR95,Flohr96b,RR96}, and the quantum
Hall effect \cite{GurFloNay97}. There have also been suggestions that
some of the so-called logarithmic operators (which appear in these
theories) might correspond to normalisable zero modes for string
backgrounds \cite{KMa95}.

By now, quite a number of such models have been analysed. They include
the WZNW model on the supergroup $GL(1,1)$ \cite{RSal92}, the $c=-2$
model \cite{Gur93,GKau96a}, gravitationally dressed conformal field
theories \cite{BKog95} and some critical disordered models
\cite{CKT95,CKT96,MSer96,CTT98}. Singular vectors of some Virasoro
models have been constructed in \cite{Flohr98}, correlation functions
have been calculated in \cite{SR,RMK}, and more structural properties
of logarithmic conformal field theories have been analysed in
\cite{Roh}. 

The new class of ``logarithmic'' chiral conformal field theories is
also of interest for more conceptual considerations of conformal field
theory. In particular, there exist logarithmic models which behave in
many respects like ordinary (non-logarithmic) chiral conformal field
theories, and it is not yet clear in which way these models differ
structurally from conventional theories. For example, there exists a
series of ``quasirational logarithmic'' Virasoro models
\cite{GKau96a}, and a series of ``rational logarithmic'' models, the
simplest of which is the triplet algebra at $c=-2$
\cite{GKau96b}. Here quasirational means that a countable set of
representations of the chiral algebra closes under fusion (with finite
fusion rules), and rational that the same holds for a certain finite
set of representations, including all (finitely many) irreducible 
representations. For these rational models Zhu's algebra \cite{Zhu} is
finite dimensional, and it should be possible to read off all
properties of the whole chiral theory from the vacuum
representation. In particular, one should be able to decide whether a
rational meromorphic conformal field theory ({\it i.e.} a chiral
algebra, for which Zhu's algebra is finite-dimensional) leads to a
logarithmic theory or not, without actually constructing all the
amplitudes. As yet little progress has been made in this direction,
although it is believed that unitary (rational) meromorphic conformal
field theories always lead to non-logarithmic theories.

The only rational logarithmic model which has been studied in detail
so far, the aforementioned triplet algebra at $c=-2$, possesses
another oddity (apart from the appearance of indecomposable reducible
representations which lead to logarithmic correlation functions), and
it is quite possible that this is true in more generality
\cite{Flohr96a}: although the theory possesses a finite fusion
algebra, the matrices corresponding to the reducible representations 
cannot be diagonalised, and a straightforward application
of Verlinde's formula does not make sense. This is mirrored
by the fact that the modular transformation properties of
some of the characters cannot be described by constant matrices
as they depend on the modular parameter $\tau$. This might
suggest that these logarithmic rational theories only
make sense as chiral theories, and that they do not correspond to
modular invariant (non-chiral) conformal field
theories. It is the purpose of this paper to demonstrate
that, at least for the case of the triplet algebra at $c=-2$, this
is not the case. The resulting theory is in every aspect a
standard (non-chiral) conformal field theory but for the property
that it does not factorise into standard chiral theories. Among
other things, this demonstrates that a non-chiral 
conformal field theory has significantly more structure than
the two chiral theories it is built from.

In the process of constructing this non-chiral theory, we shall meet a
number of novel difficulties. For example, in order to satisfy the
locality constraint, the non-chiral representation corresponding to
two indecomposable chiral representations is not simply the ordinary
tensor product of the two chiral representations, but only a certain
quotient thereof. Furthermore, in order to obtain a theory with finite
multiplicities, it will be necessary to identify states from different
(non-chiral) representations (corresponding to different
indecomposable chiral representations); this will imply that the two
different indecomposable (non-chiral) representations combine to form
a single representation with two \emph{fundamental} vectors, from
which the representation is generated by the action of the chiral
algebras. Both of these features could have been predicted from the
requirement to obtain a modular invariant theory, but it is gratifying
to see that they can also be understood as arising in this way.   
\smallskip

Our general strategy is motivated by the observation that the
fundamental objects of a (chiral) conformal field theory are the
correlation functions. In particular, as has been demonstrated in
\cite{GabGod98}, all data of a chiral conformal field theory can be
recovered from the complete set of amplitudes, and it is clear that
the same holds for non-chiral conformal field theories. In order to
construct our theory, we therefore have to determine all amplitudes
and show that they satisfy the relevant properties. In fact, it is not
actually necessary to determine \emph{all} amplitudes of the theory,
but it suffices to construct the two-, three- and four-point functions
of the \emph{fundamental} fields (the fields that correspond
to the fundamental states of the different representations). Indeed, given
the two- and three-point functions of the fundamental fields, all
other amplitudes can be derived from  these, and the consistency
conditions of all amplitudes can be reduced to those being obeyed by
the four-point functions. This reduces the problem of constructing the
theory to a finite computation which can be done in principle. 

Unfortunately, for the theory in question, some of the four-point
functions involving two or more indecomposable representations are
very complicated, and it is not feasible to calculate them
explicitly. However, since we can determine all two- and three-point
functions, all amplitudes of the theory are (in principle) determined,
and the question is only that of showing the consistency of the
resulting theory. We can then make use of the observation that the 
three-point functions (and therefore the operator product expansions)
of the fundamental fields agree with those of a certain free field
realisation of the theory. Because of the above arguments, this
implies that all amplitudes of the two theories coincide, and since
the free field theory is consistent thus establishes the consistency
of our theory.  

In the course of the construction we shall also explain how the
spectrum of the theory can be read off from the amplitudes. For the
theory in question, there exist only finitely many sectors, and each
sector appears with multiplicity one. We can then determine the
partition function of the resulting theory, and it turns out that it
is indeed modular invariant.
\medskip

The paper is organised as follows. In sections \ref{sec:corr} and
\ref{sec:ope} we describe in some detail the general strategy of our
approach. In section \ref{sec:amp} the relevant amplitudes (including
the three-point functions of the fundamental fields) are explicitly
constructed. We then explain the free field realisation in terms of
symplectic fermions in section \ref{sec:fer}, and demonstrate that the
OPEs coincide. Finally in section \ref{sec:res}, we discuss the
modular invariance of the resulting theory. We have also included a
number of appendices, where some of the more technical details of our
calculations can be found.

\section{Correlation functions}
\label{sec:corr}

Let us first explain how the correlation functions of the theory can
be determined. Let $S(w)$ be a meromorphic field of conformal weight
$h$, and denote by $S_n$ the corresponding modes
\begin{displaymath}
S(w) = \sum_n w^{n-h} S_{-n} \,.
\end{displaymath}
The action of the modes on a product of two fields is described by the
two different comultiplication formulae of \cite{Gab93}; on $N$
fields, using successively these comultiplication products, there are
$N$ different formulae which are given as
\begin{displaymath}
  \Delta^i(S_n) = S_n^{(i)} + \sum_{j\neq i} \sum_{k=-h+1}^\infty
  {n+h-1 \choose k+h-1} (z_j-z_i)^{n-k} S_k^{(j)} \,,
\end{displaymath}
where $i \in\{1, \ldots , N\}$ and $S_k^{(j)}$ is the mode $S_k$
acting on the $j$-th factor in the tensor product. (These formulae can
be obtained from the ones of \cite{Gab94} by the adjoint action of the
translation operator.)  The fusion product of these $N$ fields is
defined as the quotient space of the direct tensor product by the
relations of the form
\begin{displaymath}
e^{z_i L_{-1}}  \Delta^i(S_n)  e^{-z_i L_{-1}} =
e^{z_j L_{-1}}  \Delta^j(S_n)  e^{-z_j L_{-1}} \,.
\end{displaymath}

The correlation function $\langle\phi_1(z_1)\cdots\phi_N(z_N)\rangle$
of a given set of $N$ fields is non-trivial if the vacuum
representation is contained in the multiple fusion product of the
corresponding representations. This is equivalent to the property
that there exists a (non-trivial) linear functional $\Phi$ on the
fusion product 
$({\cal V}_1\otimes\cdots\otimes{\cal V}_N)_{\mathrm{f}}$ (where
${\cal V}_j$ is the representation corresponding to $\phi_j$ and the
subscript f denotes that this is the quotient space of the direct
tensor product) which is induced by the vacuum state (at infinity),
and which therefore satisfies 
\begin{equation}
\label{eq:deltai}
  \Phi \circ \Delta^i(S_n) = 0\,, \qquad \forall i, n < h \,.
\end{equation}
Furthermore, if there exist non-trivial null-vectors in the vacuum
representation, these will give rise to additional constraints.  The
dimension of the space of solutions for the linear functional $\Phi$
satisfying these conditions is precisely the multiplicity with which
the vacuum representation appears in the multiple fusion product of
the representations. The results of our analysis of fusion
\cite{GKau96b} determine therefore the number of different solutions. 
\medskip

To find the different correlation functions explicitly, we shall use
the following method. First we observe that any functional $\Phi$,
satisfying the above constraints, is already uniquely determined by 
its value on 
${\cal V}^{0}_1\otimes{\cal V}^{0}_2\otimes{\cal V}^{\mathrm{s}}_3
\otimes\cdots\otimes{\cal V}^{\mathrm{s}}_N$, 
where ${\cal V}^{0}$ denotes the highest weight space of a
representation ${\cal V}$, {\it i.e.} the space of states which is not
in the image of the action of the negative modes, and 
${\cal V}^{\mathrm{s}}$ denotes the special subspace \cite{Nahm94}.
Indeed, we can use the relations (\ref{eq:deltai}) to remove modes
$S_k$ with $k<-h$ from the states $\phi_i$. This allows us to
determine the value of the functional for arbitrary vectors in terms
of those on the $N$-fold tensor product of the special subspaces
${\cal V}^{\mathrm{s}}_1\otimes\cdots\otimes{\cal V}^{\mathrm{s}}_N$
\cite{Nahm94}.

We can then use the $2h-2$ relations coming from 
\begin{displaymath}
  \Phi \circ e^{z_i L_{-1}}  \Delta^i(S_n) e^{-z_i L_{-1}}   = 0\,,
\end{displaymath}
where $-h<n<h$ (in which case the formula is actually independent of
$i$), which can be written as $\Phi \circ \Delta^0(S_n) = 0$ with
\begin{equation}
\label{eq:vacrels}
  \Delta^0(S_n) = \sum_{j} \sum_{k=-h+1}^n
  {n+h-1 \choose k+h-1} z_j^{n-k} S_k^{(j)}\,.
\end{equation}
In particular, we can use these formulae to remove the negative modes
of $S$ on two of the states $\phi_1$ and $\phi_2$, say. At each stage
in the reduction process, when commuting annihilation modes through
creation modes we might again produce modes $S_k$ with $k\leq-h$ but
the sum of the conformal weights of the $n$ states decreases at each
step and, therefore, we can reduce any correlation function to a
functional on 
${\cal V}^{0}_1\otimes{\cal V}^{0}_2\otimes{\cal V}^{\mathrm{s}}_3 
\otimes\cdots\otimes{\cal V}^{\mathrm{s}}_N$. (This argument is
analogous to the argument given in \cite{Nahm94} for the case of three
representations.)
\smallskip

This argument implies that the number of different solutions is
bounded by the dimension of the space 
${\cal V}^{0}_1\otimes{\cal V}^{0}_2\otimes{\cal V}^{\mathrm{s}}_3
\otimes\cdots\otimes{\cal   V}^{\mathrm{s}}_N$. In general, the
dimension of this space need not agree with the multiplicity of the
vacuum representation in the $N$-fold fusion product, as there might
exist the analogue of the {\em spurious subspace} of Nahm
\cite{Nahm94}. Once we have found all relations, we can determine the
dependence of the correlation functions on $z_i$ by solving the system
of first order differential equations which are obtained by
identifying ${d \over dz_i} \leftrightarrow L_{-1}^{(i)}$.  
\smallskip

In practice, since we are mainly interested in the amplitudes of the
fundamental fields, we shall follow a slightly different approach. We
first solve the first order differential equations arising from the
comultiplication (\ref{eq:vacrels}), and this determines the two- and
three-point functions up to a constant. The four-point functions
depend then on an arbitrary function of the cross-ratio, and by using
the defining null vector for one of the fundamental fields
we obtain a higher order differential equation for this function which
we then solve. 
\medskip 

Non-chiral amplitudes are obtained by considering suitable linear
combinations of products of chiral amplitudes. These amplitudes are
required to be local, and this will constrain the way in which the
various chiral correlation functions can be combined. For example, for
a two-point function, locality requires that
\begin{displaymath}
  \langle\phi_1(\e^{2\pi\i}z, \e^{-2\pi\i}\bar z)\phi_2(0,0)\rangle 
  = 
  \langle\phi_1(z,\bar z)\phi_2(0,0)\rangle \,.
\end{displaymath}
On the other hand, because of (\ref{eq:vacrels}) and the property that
$L_0$ can be identified with the scaling generator of the M\"obius
group, the two-point function has to satisfy
\begin{displaymath}
  z\partial_z \langle\phi_1(z,\bar z)\phi_2(0,0)\rangle +
  \langle L_0\phi_1(z,\bar z)\phi_2(0,0)\rangle +
  \langle\phi_1(z,\bar z)L_0\phi_2(0,0)\rangle = 0 \,,
\end{displaymath}
and the analogous relation for the barred coordinate. We can
integrate this differential equation along a circle around the origin,
and find
\begin{displaymath}
  \langle\phi_1(\e^{-2\pi\i}z,\e^{2\pi\i}\bar z)\phi_2(0,0)\rangle =
  \e^{2\pi\i(h_1-\bar h_1+h_2-\bar h_2)}
  \langle \e^{2\pi\i S}\phi_1(z,\bar z)\e^{2\pi\i S}\phi_2(0,0)\rangle\,,
\end{displaymath}
where $(h_j,\bar h_j)$ are the left and right conformal weights of the
states $\phi_j$ and $S = L_0^{(\mathrm{n})} - \bar L_0^{(\mathrm{n})}$
is the nilpotent part of $L_0 - \bar L_0$. The conditions for the
two-point function to be local are thus
\begin{displaymath}
  h_1-\bar h_1+h_2-\bar h_2 \in \mathbb{Z}\,, \qquad
  \langle S^n \phi_1(z,\bar z) S^m \phi_2(0,0) \rangle = 0  \qquad  
  \forall
  n,m\in\mathbb{Z}_{\geq0}, \, m+n>0 \,.
\end{displaymath}
This has to hold for any combination of $\phi_1$ and $\phi_2$.  Since
every $N$-point function involving $\phi_1$, say, can be expanded in
terms of such two-point functions (by defining $\phi_2$ to be a
suitable contour integral of the product of the remaining $N-1$
fields), it follows that we have to have
\begin{displaymath}
  h-\bar h \in \mathbb{Z}\,, \qquad
  S \phi = 0  \,,
\end{displaymath}
where $(h,\bar h)$ are the conformal weights of any non-chiral  
field $\phi$.
\smallskip

The triplet algebra whose local conformal field theory we want to
construct has four indecomposable chiral representations which close
under fusion \cite{GKau96b}. There are two irreducible
representations, ${\cal V}_{-1/8}$, generated from an $su(2)$-singlet
state $\mu$ of weight $h=-1/8$, and ${\cal V}_{3/8}$, generated from
an $su(2)$-doublet $\nu^\alpha$ of weight $h=3/8$. The corresponding 
non-chiral irreducible representations are the ``diagonal'' tensor
products  
\begin{displaymath}
  {\cal V}_{-1/8,-1/8} = {\cal V}_{-1/8} \otimes \bar{\cal V}_{-1/8}, 
  \qquad
  {\cal V}_{3/8,3/8} = {\cal V}_{3/8} \otimes \bar{\cal V}_{3/8}
\end{displaymath}
with cyclic states $\bmu = \mu\otimes\bar\mu$ and
$\bnu^{\alpha\bar\alpha} = \nu^\alpha\otimes\bar\nu^{\bar\alpha}$, 
respectively. In this case the above constraint is manifestly
satisfied as $S\equiv 0$ on ${\cal V}_{-1/8,-1/8}$ and 
${\cal V}_{3/8,3/8}$. 

The other two representations are reducible (but indecomposable) and
are characterised by the diagrams
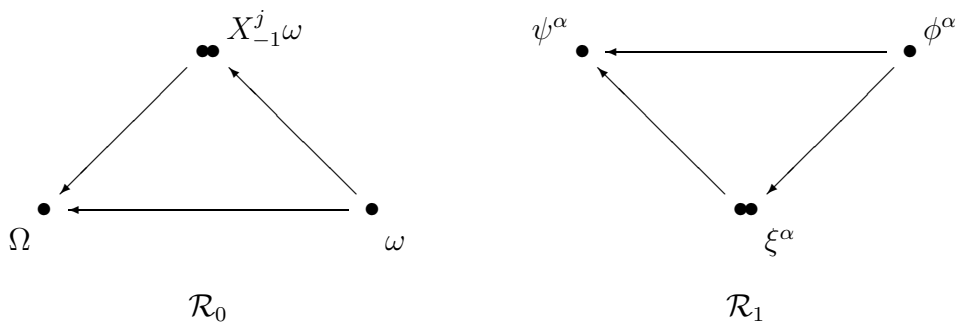
\begin{displaymath}
\begin{picture}(180,140)(-90,-50)
  \put(62,0){\vbox to 0pt
        {\vss\hbox to 0pt{\hss$\bullet$\hss}\vss}}
  \put(-62,0){\vbox to 0pt
        {\vss\hbox to 0pt{\hss$\bullet$\hss}\vss}}
  \put(2,60){\vbox to 0pt
        {\vss\hbox to 0pt{\hss$\bullet$\hss}\vss}}
  \put(-2,60){\vbox to 0pt
        {\vss\hbox to 0pt{\hss$\bullet$\hss}\vss}}
  \put(56,6){\vector(-1,1){48}}
  \put(-8,54){\vector(-1,-1){48}}
  \put(53,0){\vector(-1,0){106}}
  \put(-67,-15){\hbox to 0pt{\hss$\Omega$}}
  \put(67,-15){\hbox to 0pt{$\omega$\hss}}
  \put(7,65){\hbox to 0pt{$X^j_{-1}\omega$\hss}}
  \put(0,-40){\hbox to 0pt{\hss${\cal R}_0$\hss}}
\end{picture}
\qquad
\begin{picture}(180,140)(-90,-50)
  \put(62,60){\vbox to 0pt
        {\vss\hbox to 0pt{\hss$\bullet$\hss}\vss}}
  \put(-62,60){\vbox to 0pt
        {\vss\hbox to 0pt{\hss$\bullet$\hss}\vss}}
  \put(2,0){\vbox to 0pt
        {\vss\hbox to 0pt{\hss$\bullet$\hss}\vss}}
  \put(-2,0){\vbox to 0pt
        {\vss\hbox to 0pt{\hss$\bullet$\hss}\vss}}
  \put(56,54){\vector(-1,-1){48}}
  \put(-8,6){\vector(-1,1){48}}
  \put(53,60){\vector(-1,0){106}}
  \put(-67,65){\hbox to 0pt{\hss$\psi^\alpha$}}
  \put(67,65){\hbox to 0pt{$\phi^\alpha$\hss}}
  \put(7,-15){\hbox to 0pt{$\xi^\alpha$\hss}}
  \put(0,-40){\hbox to 0pt{\hss${\cal R}_1$\hss}}
\end{picture}
\end{displaymath}
Here, each vertex represents a representation of the chiral algebra,
and an arrow $A\longrightarrow B$ indicates that the representation
$B$ is in the image of $A$ under the action of the chiral algebra. In
the bottom row, the representations have conformal weight $h=0$, and
in the top row $h=1$. The representation ${\cal R}_0$ is generated
from a cyclic vector $\omega$ of $h=0$, which is a singlet under
$su(2)$. It forms a Jordan block for $L_0$ with $\Omega$, and the
defining relations are \cite{GKau96b}
\begin{displaymath}
  \begin{array}{rcl@{\qquad}rcl}
    L_0 \omega &=& \Omega\,, & L_0 \Omega &=& 0\,, \\
    W^a_0 \omega &=& 0\,, & W^a_0 \Omega &=& 0\,.
  \end{array}
\end{displaymath}
The four states $L_{-1}\omega$ and $W^a_{-1}\omega$, collectively
denoted by $X^j_{-1}\omega$, form two doublets under $su(2)$.

The representation ${\cal R}_1$ is generated from a doublet $\phi^\pm$
of weight $h=1$. It has two ground states $\xi^\pm$ at $h=0$ and
another doublet $\psi^\pm$ at $h=1$ forming an $L_0$ Jordan block with
$\phi^\pm$. The defining relations are \cite{GKau96b}
\begin{displaymath}
  \begin{array}{rcl@{\qquad}rcl}
    L_1 \phi^\alpha &=& -\xi^\alpha\,, &
    W^a_1 \phi^\alpha &=& t^{a\alpha}_\beta \xi^\beta\,, \\
    L_0 \phi^\alpha &=& \phi^\alpha + \psi^\alpha\,, &
    W^a_0 \phi^\alpha &=& 2t^{a\alpha}_\beta \phi^\beta\,,  
\\[\medskipamount]
    L_0 \xi^\alpha &=& 0\,, & W^a_0 \xi^\alpha &=& 0\,, \\
    L_{-1} \xi^\alpha &=& \psi^\alpha\,, &
    W^a_{-1} \xi^\alpha &=& t^{a\alpha}_\beta \psi^\beta\,, \\
    L_0 \psi^\alpha &=& \psi^\alpha\,, &
    W^a_0 \psi^\alpha &=& 2 t^{a\alpha}_\beta \psi^\beta\,.
  \end{array}
\end{displaymath}
The two states $\xi^\alpha$ form two singlets under $su(2)$ and are
thus represented by a pair of vertices in the above diagram.
\smallskip

Suppose now that we want to construct a non-chiral local field
corresponding to the tensor product of the chiral states 
$\psi \otimes \bar\psi \in {\cal R}_0\otimes \bar {\cal R}_0$.  In
order for this field to be local, the action of $S$ must vanish on
$\psi \otimes \bar\psi $. In general, however, 
$S(\psi \otimes \bar\psi)$ does not vanish in 
${\cal R}_0\otimes \bar{\cal R}_0$, and this must mean that it is only
possible to associate local fields to a certain quotient space of
${\cal R}_0\otimes\bar{\cal R}_0$. 

To determine this quotient space, we observe that $S$ commutes with
the action of both chiral algebras. This implies that every state
which is in the image space of $S$, is in the subrepresentation
generated from $S(\omega \otimes \bar\omega)$, where 
$\omega \otimes \bar\omega$ is the cyclic state for the representation
${\cal R}_0\otimes\bar{\cal R}_0$ with respect to the action of both
chiral algebras. It is therefore clear that the space by which we have
to quotient has to contain at least the subspace ${\cal N}_{0\bar0}$
which is the subrepresentation generated from 
$S(\omega \otimes \bar\omega)=\Omega\otimes\bar\omega -
\omega\otimes\bar\Omega$. For ${\cal R}_1\otimes\bar{\cal R}_1$, the
situation is analogous, and ${\cal N}_{1\bar1}$ is generated from 
$\phi^\alpha\otimes\bar\psi^{\bar\alpha}
-\psi^\alpha\otimes\bar\phi^{\bar\alpha}$. The maximal (non-chiral) 
representations which can correspond to local fields are thus of the
form
\begin{displaymath}
  {\cal R}_{0\bar0} = \left({\cal R}_0\otimes\bar{\cal R}_0\right) /
  {\cal N}_{0\bar0}\,, \qquad
  {\cal R}_{1\bar1} = \left({\cal R}_1\otimes\bar{\cal R}_1\right) /
  {\cal N}_{1\bar1}\,;
\end{displaymath}
their structure can be summarised schematically as
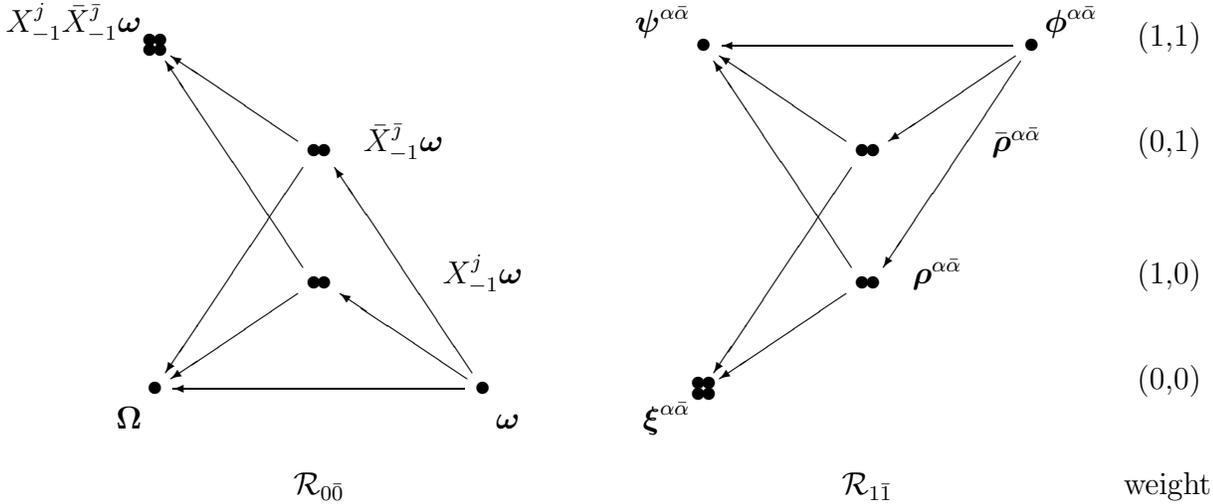
\begin{displaymath}
  \begin{picture}(184,210)(-92,-50)
    \put(2,40){\vbox to 0pt
        {\vss\hbox to 0pt{\hss$\bullet$\hss}\vss}}
    \put(2,90){\vbox to 0pt
        {\vss\hbox to 0pt{\hss$\bullet$\hss}\vss}}
    \put(62,0){\vbox to 0pt
        {\vss\hbox to 0pt{\hss$\bullet$\hss}\vss}}
    \put(56,4){\vector(-3,2){48}}
    \put(58,6){\vector(-2,3){52}}
    \put(-2,40){\vbox to 0pt
        {\vss\hbox to 0pt{\hss$\bullet$\hss}\vss}}
    \put(-2,90){\vbox to 0pt
        {\vss\hbox to 0pt{\hss$\bullet$\hss}\vss}}
    \put(-60,132){\vbox to 0pt
        {\vss\hbox to 0pt{\hss$\bullet$\hss}\vss}}
    \put(-60,128){\vbox to 0pt
        {\vss\hbox to 0pt{\hss$\bullet$\hss}\vss}}
    \put(-64,132){\vbox to 0pt
        {\vss\hbox to 0pt{\hss$\bullet$\hss}\vss}}
    \put(-64,128){\vbox to 0pt
        {\vss\hbox to 0pt{\hss$\bullet$\hss}\vss}}
    \put(-62,0){\vbox to 0pt
        {\vss\hbox to 0pt{\hss$\bullet$\hss}\vss}}
    \put(-8,36){\vector(-3,-2){48}}
    \put(-6,84){\vector(-2,-3){52}}
    \put(-8,94){\vector(-3,2){48}}
    \put(-6,46){\vector(-2,3){52}}
    \put(55,0){\vector(-1,0){110}}
    \put(-67,-15){\hbox to 0pt{\hss$\bOmega$}}
    \put(67,-15){\hbox to 0pt{$\bomega$\hss}}
    \put(-67,135){\hbox to 0pt{\hss$X^j_{-1}\bar  
X^{\bar\jmath}_{-1}\bomega$}}
    \put(47,40){\hbox to 0pt{$X^j_{-1}\bomega$\hss}}
    \put(17,90){\hbox to 0pt{$\bar X^{\bar\jmath}_{-1}\bomega$\hss}}
    \put(0,-40){\hbox to 0pt{\hss${\cal R}_{0\bar0}$\hss}}
  \end{picture}
\qquad
  \begin{picture}(184,210)(-92,-50)
    \put(2,40){\vbox to 0pt
        {\vss\hbox to 0pt{\hss$\bullet$\hss}\vss}}
    \put(2,90){\vbox to 0pt
        {\vss\hbox to 0pt{\hss$\bullet$\hss}\vss}}
    \put(62,130){\vbox to 0pt
        {\vss\hbox to 0pt{\hss$\bullet$\hss}\vss}}
    \put(56,126){\vector(-3,-2){48}}
    \put(58,124){\vector(-2,-3){52}}
    \put(-2,40){\vbox to 0pt
        {\vss\hbox to 0pt{\hss$\bullet$\hss}\vss}}
    \put(-2,90){\vbox to 0pt
        {\vss\hbox to 0pt{\hss$\bullet$\hss}\vss}}
    \put(-62,130){\vbox to 0pt
        {\vss\hbox to 0pt{\hss$\bullet$\hss}\vss}}
    \put(-60,2){\vbox to 0pt
        {\vss\hbox to 0pt{\hss$\bullet$\hss}\vss}}
    \put(-60,-2){\vbox to 0pt
        {\vss\hbox to 0pt{\hss$\bullet$\hss}\vss}}
    \put(-64,2){\vbox to 0pt
        {\vss\hbox to 0pt{\hss$\bullet$\hss}\vss}}
    \put(-64,-2){\vbox to 0pt
        {\vss\hbox to 0pt{\hss$\bullet$\hss}\vss}}
    \put(-8,36){\vector(-3,-2){48}}
    \put(-6,84){\vector(-2,-3){52}}
    \put(-8,94){\vector(-3,2){48}}
    \put(-6,46){\vector(-2,3){52}}
    \put(55,130){\vector(-1,0){110}}
    \put(-67,-15){\hbox to 0pt{\hss$\bxi^{\alpha\bar\alpha}$}}
    \put(-67,135){\hbox to 0pt{\hss$\bpsi^{\alpha\bar\alpha}$}}
    \put(67,135){\hbox to 0pt{$\bphi^{\alpha\bar\alpha}$\hss}}
    \put(17,40){\hbox to 0pt{$\brho^{\alpha\bar\alpha}$\hss}}
    \put(47,90){\hbox to 0pt{$\bar\brho^{\alpha\bar\alpha}$\hss}}
    \put(0,-40){\hbox to 0pt{\hss${\cal R}_{1\bar1}$\hss}}
  \end{picture}
  \begin{picture}(40,210)(-10,-50)
    \put(0,0){(0,0)}
    \put(0,40){(1,0)}
    \put(0,90){(0,1)}
    \put(0,130){(1,1)}
    \put(-5,-40){weight}
  \end{picture}
\end{displaymath}
Here $\bomega$ is the equivalence class of states in 
${\cal R}_{0\bar0}$ which contains as a representative 
$(\omega\otimes \bar\omega)$, and we have 
$\bOmega = L_0 \bomega =\bar L_0 \bomega$ (as $S \bomega =0$). 
Likewise, $\bphi^{\alpha\bar\alpha}$ is the equivalence class in 
${\cal R}_{1\bar1}$ with representative 
$(\phi^\alpha \otimes \bar\phi^{\bar\alpha})$. We shall always use
the convention that non-chiral states are denoted by bold symbols.
\bigskip

This solves the locality constraints for two-point functions. For
higher point amplitudes, locality will require that we combine only
suitable combinations of chiral correlation functions. This will
eliminate already a large number of possibilities, but in order to
constrain the amplitudes further, we have to turn to different
considerations relating to the spectrum of the non-chiral theory.

\section{The spectrum of the theory and the operator product expansion}
\label{sec:ope}

The considerations of the previous section only determine the (chiral)
correlation functions and the (non-chiral) amplitudes up to some
integration constants. In terms of the chiral theory alone, there is
no way of obtaining further restrictions on these constants, but for
the non-chiral theory there are further constraints which come about
as follows. The local fields of the non-chiral theory are
well-defined operators on the whole space of non-chiral states, and
are in one-to-one correspondence with these states.  (Indeed, given a
complete set of amplitudes, we can reconstruct a dense subspace of
states as the natural quotient space\footnote{We quotient by all
states whose amplitudes vanish identically.}  of the free vector
space of formal products of fields, and for this dense subspace there
exists a one-to-one correspondence between states and fields.  The
actual space of states is then defined as the weak completion, using
the natural weak topology induced by the amplitudes \cite{GabGod98}.)
This construction is by the way not possible for a chiral theory as
the chiral amplitudes carry additional labels (specifying the
different ``fusion channels").

It therefore makes sense to ask how large the space of states is, and
in particular, what the multiplicities of the various subsectors are.
This will depend crucially on the various normalisation constants of
the amplitudes, and for a generic choice, these multiplicities will
all be infinite. It is therefore an interesting question whether there
exists a special solution for the amplitudes for which the
multiplicities are finite. This is, in particular, necessary for 
a finite partition function, and only in this case can we expect to
obtain a modular invariant theory. 
\smallskip

The simplest situation (which is the one which shall be relevant in
the following) arises if all sectors of the theory appear with
multiplicity one, so that every field is uniquely characterised by its
representation properties with respect to the chiral algebra (and
there are no additional labels referring to the different copies of
the given representation). Using the action of the meromorphic fields,
every amplitude can be rewritten in terms of those that only involve
the {\it fundamental fields}, {\it i.e.} the fields whose
corresponding states are the fundamental states. Furthermore, it is 
sufficient to know the two- and three-point functions of the
fundamental fields, as these determine already the operator product
expansions (OPEs), from which higher amplitudes can be constructed by
the so-called gluing process. 

Conversely, given a set of two- and three- point functions of the
fundamental fields, we can use these to construct higher
amplitudes. In general, there are different ways of obtaining
a given $n$-point function, and a priori, it is not clear whether the
different functions so constructed agree. A sufficient (and necessary)
condition for them to agree is that they do so in the case of the
four-point functions, in which case the condition is usually referred
to as duality. (This is sufficient, as every
decomposition of the sphere into ``pairs of pants'' can be related to
any other one by a succession of ``simple moves'', where each simple
move only involves four fields \cite{MooreSeib}.) In order to
construct a theory with trivial multiplicities, it is therefore
sufficient to determine the two-, three- and four-point functions of
the fundamental fields, and show that they are local and satisfy the
duality relations; this is what we shall do in the following.
\medskip

In order to determine the $n$-point functions involving the
reducible representations from those of the irreducible ones, it is
useful to determine the operator product expansions (OPEs) of the 
fundamental fields explicitly. These (non-chiral) OPEs are effectively
built from the two chiral OPEs corresponding to the chiral
representations, and their information in turn is encoded in the
chiral fusion rules. For example, the (chiral) fusion product of
$\mu$ with itself contains only the representation ${\cal R}_0$
\cite{GKau96b} which, at lowest level, has a two-dimensional Jordan
block for $L_0$. Since $L_0$ can be identified with the scale
generator of the M\"obius transformations it follows that the chiral
OPE for $\mu(x)\mu$ has a power series expansion of the form
(compare \cite{RMK})
\begin{displaymath}
  \mu(x) \mu =
  x^{\frac14} \sum_{n=0}^\infty x^n
  \left( X_n + Y_n \ln x \right)\,,
\end{displaymath}
where $X_n, Y_n$ are states in ${\cal R}_{0}$ of weight $n$. The
states for $n>1$ are in the representation generated by those at
$n=0$, and so by acting with positive modes on them, using the
comultiplication relations, we can find recursive relations for
$X_n, Y_n$ with $n>0$ in terms of $X_0$ and $Y_0$.  It is therefore
sufficient to determine only the lowest level terms.

Without loss of generality, we can identify $X_0 = \omega$ with a
cyclic vector of ${\cal R}_0$. Then, the chiral OPE is of the form
\begin{eqnarray*}
  x^{-\frac14} \mu(x)\mu &=&
  \omega + \Omega \ln x +
  \frac12 L_{-1} \omega x +
\\*&&+
  \left(
    \frac{13}{16} L_{-1}^2 \omega - \frac38 L_{-2} \omega +
    \frac{19}{8} L_{-2}\Omega -\frac38 L_{-2}\Omega \ln x
  \right) x^2 + \cdots \,,
\end{eqnarray*}
where $\Omega = L_0 \omega$ and thus $L_0 \Omega=0$.

In a similar way we can determine the other chiral OPEs. The only
slight complication arises due to the $su(2)$ tensorial structure; for
a definition of the various $su(2)$ tensors see appendix
\ref{sec:appten}. 
\begin{eqnarray*}
  x^{\frac14} \mu(x)\nu^{\alpha} &=&
  \xi^{\alpha} +
  \left(
    \frac12 \phi^{\alpha} + \frac12 \psi^\alpha \ln x
  \right) x +
\\*&&+
  \left(
    \frac18 L_{-1} \phi^\alpha - \frac{3}{16} L_{-1} \psi^\alpha +
    \frac14 L_{-2} \xi^\alpha + \frac18 L_{-1} \psi^\alpha \ln x
  \right) x^2
  \cdots\,, \\
  x^{\frac34} \nu^{\alpha}(x)\nu^{\beta} &=&
  d^{\alpha\beta} \Bigg(
  \omega' + \Omega \ln x +
  \frac12 L_{-1} \omega' x +
\\*&&\qquad+
  \left(
    \frac{13}{16} L_{-1}^2 \omega' - \frac38 L_{-2} \omega' +
    \frac{19}{8} L_{-2}\Omega -\frac38 L_{-2}\Omega \ln x
  \right) x^2 + \cdots \Bigg)
\\*&&+
  t^{\alpha\beta}_a \left(
    4 W^a_{-1} \omega' x + W^a_{-2} x^2 +
    \cdots
  \right).
\end{eqnarray*}
The vector $\omega'$ is a cyclic vector of the representation 
${\cal R}_0$, and $\Omega = L_0 \omega'$. Here we have included a
prime for $\omega$ as the cyclic vector of the representation 
${\cal R}_0$ is not uniquely fixed. Apart from the usual freedom of
rescaling $\omega$, we can always add to a given cyclic vector
$\omega$ a multiple of $\Omega = L_0 \omega$.  (This is a special
feature of not completely reducible representations --- if the fusion
product is completely reducible, because of Schur's lemma, the only
remaining freedom is a constant for each irreducible component.) 

The (non-chiral) OPEs are obtained by multiplying the chiral
OPEs. However, if the corresponding theory is indeed local, only the
states in ${\cal R}_{j\bar\jmath}$ can occur in the OPE. To first
order in $x$ we therefore find (the expressions to second order in $x$
and $\bar x$ are listed in appendix \ref{sec:appope})
\begin{eqnarray}
\label{mumuope}
  |x|^{-\frac12} \bmu(x)\bmu &=&
  \bomega + \ln|x|^2 \bOmega +
  \cdots\,, \\
\label{munuope}
  |x|^{\frac12} \bmu(x)\bnu^{\alpha\bar\alpha} &=&
  \bxi^{\alpha\bar\alpha} +
  \frac12\brho^{\alpha\bar\alpha} x +
  \frac12\bar\brho^{\alpha\bar\alpha} \bar x +
  \frac14 \Biggl(
    \bphi^{\alpha\bar\alpha} +
    \ln|x|^2 \bpsi^{\alpha\bar\alpha} 
  \Biggr) |x|^2 +
  \cdots\,, \\
\label{nunuope}
  |x|^{\frac32} \bnu^{\alpha\bar\alpha}(x)\bnu^{\beta\bar\beta} &=&
  -\frac14 d^{\alpha\beta} d^{\bar\alpha\bar\beta}
  \Biggl( 
    \bomega' + (\ln|x|^2 + 4) \bOmega' + \cdots
  \Biggr)\,.
\end{eqnarray}
In the last OPE we have made use of the freedom to redefine $\bomega'$
and $\bOmega'$ for later convenience. We can regard these OPEs as 
{\it defining} the states $\bomega, \bOmega$ and $\bomega', \bOmega'$,
and it is therefore not clear whether  $\bomega= \bomega'$ and
$\bOmega=\bOmega'$. Indeed, we do not yet know whether all amplitudes
involving $\bomega$ agree with those involving $\bomega'$, and
similarly for $\bOmega$ and $\bOmega'$.

\section{Amplitudes}
\label{sec:amp}

In this section we shall construct the relevant amplitudes
explicitly. As a first step we fix the normalisation of the two-point 
functions of the irreducible fundamental fields
\begin{eqnarray*}
  \langle\bmu(z_1)\bmu(z_2)\rangle &=& 
  {\cal D} |z_{12}|^{1/2}\,, 
  \\
  \langle\bnu^{\alpha\bar\alpha}(z_1)\bnu^{\beta\bar\beta}(z_2)\rangle
  &=& 
  -\frac14 {\cal D} d^{\alpha\beta} d^{\bar\alpha\bar\beta}
  |z_{12}|^{-3/2}\,,
\end{eqnarray*}
where $z_{12} = z_1 - z_2$. We have also included a factor of $-1/4$
for later convenience. We can then use the OPEs
(\ref{mumuope},\,\ref{nunuope}) to deduce the one-point functions of
the reducible representations 
\begin{displaymath}
  \langle\bOmega\rangle = 
  \langle\bOmega'\rangle = 
  \langle\bxi^{\alpha\bar\alpha}\rangle = 0\,, 
  \qquad
  \langle\bomega\rangle = {\cal D}\,, 
  \qquad
  \langle\bomega'\rangle = {\cal D}\,. 
\end{displaymath}
\medskip

Next, we calculate the chiral four-point functions involving
four irreducible fields, following the strategy of section
\ref{sec:corr}. We then form the most general linear combination of
two such four-point functions which give rise to local amplitudes
\begin{eqnarray*}
  \langle\bmu(z_1)\bmu(z_2)\bmu(z_3)\bmu(z_4)\rangle 
  & = &
  -\pi {\cal C}_0
  \left|\frac{z_{12}z_{34}z_{14}z_{23}}{z_{13}z_{24}}\right|^{\frac12}
  \left[
    K(x)\tilde K(\bar x) + \tilde K(x)K(\bar x) \right] 
  \,, \\
  \langle\bmu(z_1)\bmu(z_2)\bmu(z_3)\bnu^{\alpha\bar\alpha}(z_4)\rangle
  & = & 
  -{\cal B}_0^{\alpha\bar\alpha}
  \left|\frac{z_{12}z_{13}z_{23}}{z_{14}z_{24}z_{34}}\right|^{\frac12}
  \,,  \\
  \langle\bmu(z_1)\bmu(z_2)\bnu^{\alpha\bar\alpha}(z_3)
  \bnu^{\beta\bar\beta}(z_4)\rangle 
  & = & 
  \frac{\pi}{4} {\cal C}_1 d^{\alpha\beta} d^{\bar\alpha\bar\beta}
  \left|
    \frac{z_{12}z_{13}z_{24}}{z_{14}z_{23}z_{34}^3}
  \right|^{\frac12}
  \left[
    E(x)\tilde D_2(\bar x) + \tilde D_2(x)E(\bar x)
  \right]
  \,, \\
  \langle\bmu(z_1)\bnu^{\alpha\bar\alpha}(z_2)
  \bnu^{\beta\bar\beta}(z_3)
  \bnu^{\gamma\bar\gamma}(z_4)\rangle 
  & = & 
  \frac14 {\cal B}_1^{\eta\bar\eta}
  \left|\frac{z_{24}}{z_{12}z_{14}}\right|^{\frac12}
  \left|\frac{z_{13}}{z_{23}z_{34}}\right|^{\frac32}
  \times 
  \\*&&
  \biggl(
    h^{\beta\gamma\alpha}_\eta
    h^{\bar\beta\bar\gamma\bar\alpha}_{\bar\eta}
    |x|^2
    -
    h^{\beta\gamma\alpha}_\eta
    d^{\bar\beta\bar\gamma} \delta^{\bar\alpha}_{\bar\eta}
    \frac12 x(2-\bar x) 
    \\*&&\quad
    -
    d^{\beta\gamma} \delta^{\alpha}_{\eta}
    h^{\bar\beta\bar\gamma\bar\alpha}_{\bar\eta}
    \frac12 (2-x)\bar x
    +
    d^{\beta\gamma} \delta^{\alpha}_{\eta}
    d^{\bar\beta\bar\gamma} \delta^{\bar\alpha}_{\bar\eta}
    \frac14 |2-x|^2
  \bigg)
  \,, \\
  \langle\bnu^{\alpha\bar\alpha}(z_1)\bnu^{\beta\bar\beta}(z_2)
  \bnu^{\gamma\bar\gamma}(z_3)\bnu^{\delta\bar\delta}(z_4)\rangle & = &
  \frac{\pi}{16} {\cal C}_2
  \left|\frac{z_{13}z_{24}}{z_{12}z_{34}z_{14}z_{23}}\right|^{\frac32}
  \times
  \\*&&
  \bigg(
    h^{\alpha\beta\gamma\delta}
    h^{\bar\alpha\bar\beta\bar\gamma\bar\delta}
    \left[
      F_1(x)\tilde F_1(\bar x) + \tilde F_1(x) F_1(\bar x)
    \right] 
    \\*&& \quad
    +
    h^{\alpha\beta\gamma\delta}
    d^{\bar\alpha\bar\beta} d^{\bar\gamma\bar\delta}
    \left[
      F_1(x)\tilde F_2(\bar x) + \tilde F_1(x) F_2(\bar x)
    \right] 
    \\*&& \quad
    +
    d^{\alpha\beta} d^{\gamma\delta}
    h^{\bar\alpha\bar\beta\bar\gamma\bar\delta}
    \left[
      F_2(x)\tilde F_1(\bar x) + \tilde F_2(x) F_1(\bar x)
    \right] 
    \\*&& \quad
    +
    d^{\alpha\beta} d^{\gamma\delta}
    d^{\bar\alpha\bar\beta} d^{\bar\gamma\bar\delta}
    \left[
      F_2(x)\tilde F_2(\bar x) + \tilde F_2(x) F_2(\bar x)
    \right]
  \bigg)
  \,.
\end{eqnarray*}
The functions $K, E, D_2, F_1, F_2$ and their $(\tilde\cdot)$
counterparts are related to complete elliptic integrals (see appendix
\ref{sec:appell}). ${\cal C}_0, {\cal C}_1, {\cal C}_2, 
{\cal B}_0^{\alpha\bar\alpha}$ and ${\cal B}_1^{\alpha\bar\alpha}$ are
(at this stage) arbitrary constants. The factors of $\pi$ were
introduced for later convenience.
The cross-ratio $x$ is defined by $x=(z_{12}z_{34})/(z_{13}z_{24})$,
where as always from now on $z_{ij}\equiv z_i - z_j$. The above
expressions are well-defined for $x$ near the origin. Using the
formulae of appendix \ref{sec:appell}, we can analytically continue
the amplitudes to different values for $x$ which can then be written
in terms of amplitudes where the fields are in a different order (and
the corresponding cross-ratio is again small); the relevant
expressions are given in appendix \ref{sec:appamp}.

Next, we use the OPEs (\ref{mumuope} -- \ref{nunuope})
to determine the non-chiral three-point functions of two irreducible 
representations and one reducible representation. In general, there
are different limits of a given four-point amplitude that should give
rise to the same three-point amplitude, and the duality relations
require that all these three-point functions are indeed the
same. For the situation in question, this follows manifestly from the
expressions above and the ones of appendix \ref{sec:appamp}.

Let us consider first the states of weight $(0,0)$ in the reducible
representation. 
\begin{eqnarray*}
  \langle\bmu(z_1)\bmu(z_2)\bomega(z_3)\rangle &=&
  -{\cal C}_0 |z_{12}|^{\frac12} \left(
    8\ln2 + \ln\left|\frac{z_{13}z_{23}}{z_{12}}\right|^2
  \right), \\
  \langle\bmu(z_1)\bmu(z_2)\bomega'(z_3)\rangle &=&
  - {\cal C}_1 |z_{12}|^{\frac12} \left(
    8\ln2 + \ln\left|\frac{z_{13}z_{23}}{z_{12}}\right|^2
  \right), \\
  \langle\bmu(z_1)\bmu(z_2)\bOmega(z_3)\rangle &=&
  {\cal C}_0 |z_{12}|^{\frac12}, \\
  \langle\bmu(z_1)\bmu(z_2)\bOmega'(z_3)\rangle &=&
  {\cal C}_1 |z_{12}|^{\frac12}, \\
  \langle\bmu(z_1)\bmu(z_2)\bxi^{\gamma\bar\gamma}(z_3)\rangle &=&
  -{\cal B}_0^{\gamma\bar\gamma} |z_{12}|^{\frac12}, \\[\medskipamount]
  \langle\bmu(z_1)\bnu^{\beta\bar\beta}(z_2)\bomega(z_3)\rangle &=&
  -{\cal B}_0^{\beta\bar\beta} |z_{12}|^{-\frac12} |z_{13}|
  |z_{23}|^{-1}, \\* 
  \langle\bmu(z_1)\bnu^{\beta\bar\beta}(z_2)\bomega'(z_3)\rangle &=&
  -{\cal B}_1^{\beta\bar\beta} |z_{12}|^{-\frac12} |z_{13}|
  |z_{23}|^{-1}, \\* 
  \langle\bmu(z_1)\bnu^{\beta\bar\beta}(z_2)\bOmega(z_3)\rangle &=& 0,
  \\
  \langle\bmu(z_1)\bnu^{\beta\bar\beta}(z_2)\bOmega'(z_3)\rangle &=& 0,
  \\
  \langle\bmu(z_1)\bnu^{\beta\bar\beta}(z_2)
  \bxi^{\gamma\bar\gamma}(z_3)\rangle &=& 0, \\[\medskipamount]
  \langle\bnu^{\alpha\bar\alpha}(z_1)\bnu^{\beta\bar\beta}(z_2)
  \bomega(z_3)\rangle &=&
  \frac14 {\cal C}_1 d^{\alpha\beta} d^{\bar\alpha\bar\beta}
  |z_{12}|^{-\frac32} \left(
    8\ln2 - 4 + \ln\left|\frac{z_{13}z_{23}}{z_{12}}\right|^2
  \right), \\
  \langle\bnu^{\alpha\bar\alpha}(z_{13})\bnu^{\beta\bar\beta}(z_2)
  \bomega'(z_3)\rangle &=&
  \frac14 {\cal C}_2 d^{\alpha\beta} d^{\bar\alpha\bar\beta}
  |z_{12}|^{-\frac32} \left(
    8\ln2 - 4 + \ln\left|\frac{z_{13}z_{23}}{z_{12}}\right|^2
  \right), \\
  \langle\bnu^{\alpha\bar\alpha}(z_1)\bnu^{\beta\bar\beta}(z_2)
  \bOmega(z_3)\rangle &=&
  -\frac14 {\cal C}_1 d^{\alpha\beta} d^{\bar\alpha\bar\beta}
  |z_{12}|^{-\frac32}, \\
  \langle\bnu^{\alpha\bar\alpha}(z_1)\bnu^{\beta\bar\beta}(z_2)
  \bOmega'(z_3)\rangle &=&
  -\frac14 {\cal C}_2 d^{\alpha\beta} d^{\bar\alpha\bar\beta}
  |z_{12}|^{-\frac32}, \\
  \langle\bnu^{\alpha\bar\alpha}(z_1)\bnu^{\beta\bar\beta}(z_2)
  \bxi^{\gamma\bar\gamma}(z_3)\rangle &=&
  \frac14 
  {\cal B}_1^{\gamma\bar\gamma}  d^{\alpha\beta} d^{\bar\alpha\bar\beta}
  |z_{12}|^{-\frac32}.
\end{eqnarray*}
The fields $\bomega'$ and $\bOmega'$ have the same functional
behaviour as $\bomega$ and $\bOmega$, respectively, and only the
relative normalisation of the three-point functions differs. This was
to be expected since they correspond to isomorphic
representations. However, from the point of view of the non-chiral
theory, the state $\bomega$ (and likewise for $\bOmega$, {\it etc.})
will appear with non-trivial multiplicity unless
\begin{equation}
\label{eq:l00-cond}
  \frac{{\cal C}_1}{{\cal C}_0} = \frac{{\cal C}_2}{{\cal C}_1} =
  \frac{{\cal B}_1^{\alpha\bar\alpha}}{{\cal B}_0^{\alpha\bar\alpha}}
  \equiv 1\,, 
\end{equation}
in which case $\bomega' = \bomega, \bOmega' = \bOmega$. Since we want
to construct a theory with trivial multiplicities, we therefore choose
\begin{equation}
\label{eq:parm}
  {\cal C}_2 = {\cal C}_1 = {\cal C}_0\,, 
  \qquad
  {\cal B}_0^{\alpha\bar\alpha} = {\cal B}_1 ^{\alpha\bar\alpha} =
  \Theta^{\alpha\bar\alpha} {\cal C}_0\,.
\end{equation}
Apart from the identification of ${\cal R}'_{0,\bar0}$ with 
${\cal R}_{0,\bar0}$, the results also imply
\begin{displaymath}
  \bxi^{\alpha\bar\alpha} = -\Theta^{\alpha\bar\alpha} \bOmega\,.
\end{displaymath}
\medskip

At level $(1,1)$ of the reducible representation we find
\begin{eqnarray*}
  \langle\bmu(z_1)\bmu(z_2)
  X^j_{-1}\bar X^{\bar\jmath}_{-1}\bomega(z_3)\rangle &=& 0\,, \\
  \langle\bmu(z_1)\bmu(z_2)\bpsi^{\gamma\bar\gamma}(z_3)\rangle &=&
  0\,, \\
  \langle\bmu(z_1)\bnu^{\beta\bar\beta}(z_2)
  X^j_{-1}\bar X^{\bar\jmath}_{-1}\bomega(z_3)\rangle &=&
  - \frac14 b^{j\beta}_\eta b^{\bar\jmath\bar\beta}_{\bar\eta}
  \Theta^{\eta\bar\eta} {\cal C}_0
  |z_{12}|^{\frac32} |z_{13}|^{-1} |z_{23}|^{-3}\,, \\*
  \langle\bmu(z_1)\bnu^{\beta\bar\beta}(z_2)
  \bpsi^{\gamma\bar\gamma}(z_3)\rangle &=&
  - \frac14 d^{\beta\gamma} d^{\bar\beta\bar\gamma} {\cal C}_0 
  |z_{12}|^{\frac32} |z_{13}|^{-1} |z_{23}|^{-3}\,, \\
  \langle\bnu^{\alpha\bar\alpha}(z_1)\bnu^{\beta\bar\beta}(z_2)
  X^j_{-1}\bar X^{\bar\jmath}_{-1}\bomega(z_3)\rangle &=& 0\,, \\
  \langle\bnu^{\alpha\bar\alpha}(z_1)\bnu^{\beta\bar\beta}(z_2)
  \bpsi^{\gamma\bar\gamma}(z_3)\rangle &=& 0\,,
\end{eqnarray*}
where
\begin{displaymath}
  b^{j\alpha}_\beta = (\delta^\alpha_\beta, -t^{a\alpha}_\beta)\,,
  \qquad\mbox{for $X^j = (L, W^a)$}\,.
\end{displaymath}
The three-point functions with $\bomega'$ are the same as for
$\bomega$. Furthermore, we note that we can identify (at least on the
level of the above amplitudes)
\begin{displaymath}
  X^j_{-1}\bar X^{\bar\jmath}_{-1}\bomega =
  b^j_{\beta\gamma} b^{\bar\jmath}_{\bar\beta\bar\gamma} 
  \Theta^{\beta\bar\beta}
  \bpsi^{\gamma\bar\gamma},
\end{displaymath}
where $b^j_{\beta\gamma} = b^{j\eta}_\beta d_{\eta\gamma}$. The other 
states at level $(1,1)$ are $\bphi^{\gamma\bar\gamma}$ with
three-point functions
\begin{eqnarray*}
  \langle\bmu(z_1)\bmu(z_2)\bphi^{\gamma\bar\gamma}(z_3)\rangle &=&
  - \frac14 \Theta^{\gamma\bar\gamma} {\cal C}_0 |z_{12}|^{\frac12}
  \left|\frac{z_{13}+z_{23}}{z_{13}z_{23}}\right|^2\,, \\ 
  \langle\bmu(z_1)\bnu^{\beta\bar\beta}(z_2)
  \bphi^{\gamma\bar\gamma}(z_3)\rangle &=&
  \frac14 d^{\beta\gamma} d^{\bar\beta\bar\gamma} {\cal C}_0 
  |z_{12}|^{\frac32} |z_{13}|^{-1} |z_{23}|^{-3} \times
  \nonumber\\*&&
  \left( 8\ln2 + \ln\left|\frac{z_{13}z_{23}}{z_{12}}\right|^2 +
  2\frac{z_{13}+z_{23}}{z_{12}} + 
  2\frac{\bar z_{13}+\bar z_{23}}{\bar z_{12}}\right)\,,
  \\* 
  \langle\bnu^{\alpha\bar\alpha}(z_1)\bnu^{\beta\bar\beta}(z_2)
  \bphi^{\gamma\bar\gamma}(z_3)\rangle &=&
  \frac14 {\cal C}_0 
  |z_{12}|^{-\frac32} |z_{13}z_{23}|^{-1}
  \times \\*&&
  \Bigg( \frac14 d^{\alpha\beta} d^{\bar\alpha\bar\beta} 
    \Theta^{\gamma\bar\gamma} 
    |z_{13}+z_{23}|^2
    +
    d^{\alpha\beta} h^{\bar\alpha\bar\beta\bar\gamma}_{\bar\eta} 
    \Theta^{\gamma\bar\eta} 
    (z_{13}+z_{23}) \bar z_{12} 
    \\*&&\quad+
    h^{\alpha\beta\gamma}_{\eta} d^{\bar\alpha\bar\beta} 
    \Theta^{\eta\bar\gamma} 
    z_{12} (\bar z_{13}+\bar z_{23}) +
    4 h^{\alpha\beta\gamma}_{\eta}
    h^{\bar\alpha\bar\beta\bar\gamma}_{\bar\eta} 
    \Theta^{\eta\bar\eta} 
    |z_{12}|^2
  \Bigg)\,.
\end{eqnarray*}
Our choice of the normalisation constants (\ref{eq:parm}) so far
guarantees that the irreducible subrepresentations of
${\cal R}_{0\bar0}$ and ${\cal R}_{1\bar1}$ (which are at level
$(0,0)$ and $(1,1)$) only have trivial multiplicity in
the non-chiral theory. Let us analyse now how many states appear at 
level $(1,0)$. Using the OPEs
(\ref{mumuope} -- \ref{nunuope}) together with the
four-point functions of the 
irreducible representations we find
\begin{eqnarray*}
  \langle\bmu(z_1)\bmu(z_2)L_{-1}\bomega(z_3)\rangle &=&
  {\cal C}_0 |z_{12}|^{\frac12}
  \frac{z_{13}+z_{23}}{z_{13}z_{23}}\,, \\*
  \langle\bmu(z_1)\bmu(z_2)W^a_{-1}\bomega(z_3)\rangle &=& 0\,, \\*
  \langle\bmu(z_1)\bmu(z_2)\brho^{\gamma\bar\gamma}(z_3)\rangle &=&
  \frac12 \Theta^{\gamma\bar\gamma} {\cal C}_0 
  |z_{12}|^{\frac12} \frac{z_{13}+z_{23}}{z_{13}z_{23}}\,, \\
  \langle W^a_{-2}\bmu(z_1)\bmu(z_2)L_{-1}\bomega(z_3)\rangle &=& 0\,, \\*
  \langle W^a_{-2}\bmu(z_1)\bmu(z_2)W^b_{-1}\bomega(z_3)\rangle &=&
  \frac12 g^{ab} {\cal C}_0 |z_{12}|^{\frac12}
  z_{13}^{-3} z_{23} z_{12}^{-1}\,, \\*
  \langle W^a_{-2}\bmu(z_1)\bmu(z_2)
  \brho^{\gamma\bar\gamma}(z_3)\rangle &=&
  t^{a\gamma}_\eta \Theta^{\eta\bar\gamma} {\cal C}_0 
  |z_{12}|^{\frac12}
  z_{13}^{-3} z_{23} z_{12}^{-1}\,, \\
  \langle\bmu(z_1)\bnu^{\alpha\bar\alpha}(z_2)
  X^j_{-1}\bomega(z_3)\rangle &=&
  -\frac12 b^{j\alpha}_\beta \Theta^{\beta\bar\alpha} {\cal C}_0 
  |z_{12}|^{-\frac12} |z_{13}| |z_{23}|^{-1}
  \frac{z_{12}}{z_{13}z_{23}}\,, \\*
  \langle\bmu(z_1)\bnu^{\alpha\bar\alpha}(z_2)
  \brho^{\gamma\bar\gamma}(z_3)\rangle &=&
  -\frac12 d^{\alpha\gamma} d^{\bar\alpha\bar\gamma} {\cal C}_0 
  |z_{12}|^{-\frac12} |z_{13}| |z_{23}|^{-1} 
  \frac{z_{12}}{z_{13}z_{23}}\,, \\
  \langle\bnu^{\alpha\bar\alpha}(z_1)\bnu^{\beta\bar\beta}(z_2)
  L_{-1}\bomega(z_3)\rangle &=&
  -\frac14 d^{\alpha\beta} d^{\bar\alpha\bar\beta} {\cal C}_0
  |z_{12}|^{-\frac32} \frac{z_{13}+z_{23}}{z_{13}z_{23}}\,, \\
  \langle\bnu^{\alpha\bar\alpha}(z_1)\bnu^{\beta\bar\beta}(z_2)
  W^a_{-1}\bomega(z_3)\rangle &=&
  -\frac12 t^{a\alpha\beta} d^{\bar\alpha\bar\beta} {\cal C}_0
  |z_{12}|^{-\frac32} \frac{z_{12}}{z_{13}z_{23}}\,, \\
  \langle\bnu^{\alpha\bar\alpha}(z_1)\bnu^{\beta\bar\beta}(z_2)
  \brho^{\gamma\bar\gamma}(z_3)\rangle &=&
  -\frac18 d^{\alpha\beta} d^{\bar\alpha\bar\beta} 
  \Theta^{\gamma\bar\gamma} {\cal C}_0 
  |z_{12}|^{-\frac32} \frac{z_{13}+z_{23}}{z_{13}z_{23}}
  \\*&&{}
  -\frac12 h^{\alpha\beta\gamma}_\eta d^{\bar\alpha\bar\beta}  
  \Theta^{\eta\bar\gamma} {\cal C}_0 
  |z_{12}|^{-\frac32} \frac{z_{12}}{z_{13}z_{23}}\,. \\
\end{eqnarray*}
We observe that the states $X^j_{-1}\bomega$ and
$\brho^{\gamma\bar\gamma}$ are linearly dependent provided that  
\begin{equation}
\label{eq:l10-cond}
  \det \Theta = 1\,.
\end{equation}
As we shall show below this constraint is also required by the
consistency of the amplitude $\langle\bmu\bmu\bomega\bomega\rangle$.
We therefore have
\begin{displaymath}
  \brho^{\gamma\bar\gamma} = 
    \frac12 \Theta^{\gamma\bar\gamma} L_{-1}\bomega +
    2 t^\gamma_{a\alpha} \Theta^{\alpha\bar\gamma} W^a_{-1}\bomega
\end{displaymath}
and
\begin{displaymath}
  L_{-1}\bomega = \tilde\Theta_{\bar\alpha\alpha}
  \brho^{\alpha\bar\alpha}, \qquad
  W^a_{-1}\bomega = t^{a\alpha}_\beta \tilde\Theta_{\bar\alpha\alpha}
  \brho^{\beta\bar\alpha}\,,
\end{displaymath}
where $t^\gamma_{a\alpha} = g_{ab} t^{b\gamma}_\alpha$ and $\tilde\Theta$
is the inverse of $\Theta$,
\begin{displaymath}
  \tilde \Theta_{\bar\alpha\alpha} = \frac{1}{\det\Theta}
  \pmatrix{ \Theta^{--} & -\Theta^{+-} \cr -\Theta^{-+} & \Theta^{++}},
  \qquad
  \Theta^{\alpha\bar\alpha} \tilde \Theta_{\bar\alpha\beta} =
  \delta^\alpha_\beta\,,
  \qquad
  \tilde \Theta_{\bar\alpha\alpha} \Theta^{\alpha\bar\beta} =
  \delta_{\bar\alpha}^{\bar\beta}\,.
\end{displaymath}
In the same way we can express $\bar\brho^{\gamma\bar\gamma}$ in terms
of $\bar X_{-1}^{\bar\jmath}\bomega$ and vice versa.  
This implies that the two reducible representations combine to form
one representations ${\cal R}$ whose structure is summarised in the 
following diagram:
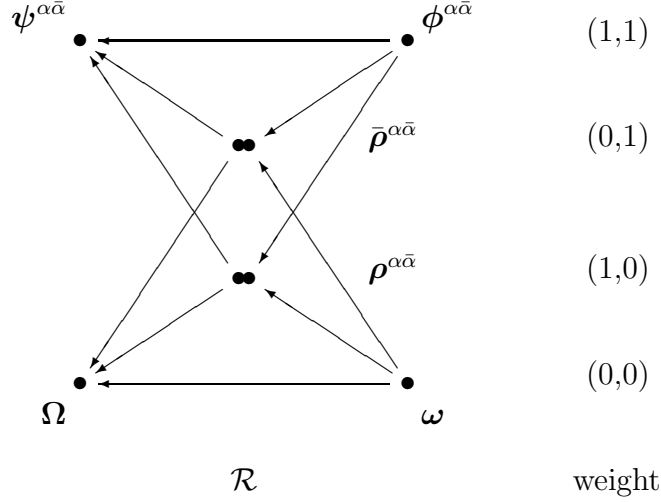
\begin{center}
  \begin{picture}(184,210)(-92,-50)
    \put(2,40){\vbox to 0pt
        {\vss\hbox to 0pt{\hss$\bullet$\hss}\vss}}
    \put(2,90){\vbox to 0pt
        {\vss\hbox to 0pt{\hss$\bullet$\hss}\vss}}
    \put(62,130){\vbox to 0pt
        {\vss\hbox to 0pt{\hss$\bullet$\hss}\vss}}
    \put(62,0){\vbox to 0pt
        {\vss\hbox to 0pt{\hss$\bullet$\hss}\vss}}
    \put(56,4){\vector(-3,2){48}}
    \put(58,6){\vector(-2,3){52}}
    \put(56,126){\vector(-3,-2){48}}
    \put(58,124){\vector(-2,-3){52}}
    \put(-2,40){\vbox to 0pt
        {\vss\hbox to 0pt{\hss$\bullet$\hss}\vss}}
    \put(-2,90){\vbox to 0pt
        {\vss\hbox to 0pt{\hss$\bullet$\hss}\vss}}
    \put(-62,130){\vbox to 0pt
        {\vss\hbox to 0pt{\hss$\bullet$\hss}\vss}}
    \put(-62,0){\vbox to 0pt
        {\vss\hbox to 0pt{\hss$\bullet$\hss}\vss}}
    \put(-8,36){\vector(-3,-2){48}}
    \put(-6,84){\vector(-2,-3){52}}
    \put(-8,94){\vector(-3,2){48}}
    \put(-6,46){\vector(-2,3){52}}
    \put(55,0){\vector(-1,0){110}}
    \put(55,130){\vector(-1,0){110}}
    \put(-67,-15){\hbox to 0pt{\hss$\bOmega$}}
    \put(67,-15){\hbox to 0pt{$\bomega$\hss}}
    \put(-67,135){\hbox to 0pt{\hss$\bpsi^{\alpha\bar\alpha}$}}
    \put(67,135){\hbox to 0pt{$\bphi^{\alpha\bar\alpha}$\hss}}
    \put(47,40){\hbox to 0pt{$\brho^{\alpha\bar\alpha}$\hss}}
    \put(47,90){\hbox to 0pt{$\bar\brho^{\alpha\bar\alpha}$\hss}}
    \put(0,-40){\hbox to 0pt{\hss${\cal R}$\hss}}
  \end{picture}
\qquad
  \begin{picture}(40,210)(-10,-50)
    \put(0,0){(0,0)}
    \put(0,40){(1,0)}
    \put(0,90){(0,1)}
    \put(0,130){(1,1)}
    \put(-5,-40){weight}
  \end{picture}
\end{center}
The representation ${\cal R}$ does not have one cyclic state, but
instead is generated from a state $\bomega$ of weight $(0,0)$ and
the four states $\bphi^{\alpha\bar\alpha}$ of weight $(1,1)$. The
non-trivial defining relations are
\begin{displaymath}
  \begin{array}{rcl@{\qquad}rcl}
    L_0 \bomega &=& \bOmega, & W^a_0 \bomega &=& 0, \\ 
    L_0 \bOmega &=& 0, & W^a_0 \bOmega &=& 0, \\ 
    L_{-1} \bomega &=& \tilde\Theta_{\bar\alpha\alpha}
    \brho^{\alpha\bar\alpha}, & 
    W^a_{-1} \bomega &=& t^{a\alpha}_\beta \tilde\Theta_{\bar\alpha\alpha}
    \brho^{\beta\bar\alpha}, \\ [\bigskipamount]
    L_0 \brho^{\alpha\bar\alpha} &=& 
    \brho^{\alpha\bar\alpha}, & 
    W^a_0 \brho^{\alpha\bar\alpha} &=& 
    2t^{a\alpha}_\beta \brho^{\beta\bar\alpha}, \\ 
    L_1 \brho^{\alpha\bar\alpha} &=& 
    \Theta^{\alpha\bar\alpha} \, \bOmega, & 
    W^a_1 \brho^{\alpha\bar\alpha} &=& 
    t^{a\alpha}_\beta \Theta^{\beta\bar\alpha} \, \bOmega, \\ 
    L_{-1} \bar\brho^{\alpha\bar\alpha} &=& 
    \bpsi^{\alpha\bar\alpha}, & 
    W^a_{-1} \bar\brho^{\alpha\bar\alpha} &=& 
    t^{a\alpha}_\beta \bpsi^{\beta\bar\alpha}, \\[\bigskipamount]
    L_0 \bphi^{\alpha\bar\alpha} &=& 
    \bphi^{\alpha\bar\alpha} + \bpsi^{\alpha\bar\alpha}, & 
    W^a_0 \bphi^{\alpha\bar\alpha} &=& 
    2 t^{a\alpha}_\beta \bphi^{\beta\bar\alpha}, \\ 
    L_0 \bpsi^{\alpha\bar\alpha} &=& 
    \bpsi^{\alpha\bar\alpha}, & 
    W^a_0 \bpsi^{\alpha\bar\alpha} &=& 
    2 t^{a\alpha}_\beta \bpsi^{\beta\bar\alpha}, \\ 
    L_1 \bphi^{\alpha\bar\alpha} &=& 
    - \bar\brho^{\alpha\bar\alpha}, & 
    W^a_1 \bphi^{\alpha\bar\alpha} &=& 
    -t^{a\alpha}_\beta \bar\brho^{\beta\bar\alpha}, 
  \end{array}
\end{displaymath}
together with their anti-chiral counterparts. 

Given the above three-point functions we can now apply the OPEs 
(\ref{mumuope} -- \ref{nunuope}) again, and deduce the two-point
functions for the reducible representations ${\cal R}$. We find
\begin{eqnarray*}
  \langle\bomega\bomega\rangle &=& 
  -{\cal C}_0 \left( 8\ln2 + 2\ln|z_{12}|^2\right)\,,
  \\
  \langle\bomega\bOmega\rangle &=& {\cal C}_0\,, \\
  \langle\bOmega\bOmega\rangle &=& 0\,,
  \\
  \langle\bomega\brho^{\beta\bar\beta}\rangle &=&
  \Theta^{\beta\bar\beta} {\cal C}_0 z_{12}^{-1}\,, \\
  \langle\bomega\bar\brho^{\beta\bar\beta}\rangle &=&
  \Theta^{\beta\bar\beta} {\cal C}_0 \bar z_{12}^{-1}\,, \\
  \langle\bomega\bphi^{\beta\bar\beta}\rangle &=&
  -\Theta^{\beta\bar\beta} {\cal C}_0 |z_{12}|^{-2}\,,
  \\
  \langle\brho^{\alpha\bar\alpha}\brho^{\beta\bar\beta}\rangle &=&
  -d^{\alpha\beta} d^{\bar\alpha\bar\beta} {\cal C}_0 
  z_{12}^{-2}\,, \\
  \langle\brho^{\alpha\bar\alpha}\bar\brho^{\beta\bar\beta}\rangle &=&
  0\,, \\
  \langle\brho^{\alpha\bar\alpha}\bphi^{\beta\bar\beta}\rangle &=&
  d^{\alpha\beta} d^{\bar\alpha\bar\beta} {\cal C}_0 
  z_{12}^{-2} \bar z_{12}^{-1}\,,
  \\
  \langle\brho^{\alpha\bar\alpha}\bpsi^{\beta\bar\beta}\rangle &=& 0\,,
  \\
  \langle\bphi^{\alpha\bar\alpha}\bphi^{\beta\bar\beta}\rangle &=&
  d^{\alpha\beta} d^{\bar\alpha\bar\beta} {\cal C}_0 
  |z_{12}|^{-4}
  \left( 8\ln2 + 2 + 2\ln|z_{12}|^2 \right) \,,
  \\
  \langle\bphi^{\alpha\bar\alpha}\bpsi^{\beta\bar\beta}\rangle &=&
  -d^{\alpha\beta} d^{\bar\alpha\bar\beta} {\cal C}_0 
  |z_{12}|^{-4}\,,
  \\
  \langle\bpsi^{\alpha\bar\alpha}\bpsi^{\beta\bar\beta}\rangle &=&
  0\,.
\end{eqnarray*}
We can also determine further OPEs from the three-point functions
involving one reducible and two irreducible fields. For example
it follows from the amplitudes $\langle\bmu\bmu\bomega\rangle$ and
$\langle\bmu\bnu^{\alpha\bar\alpha}\bomega\rangle$ that the OPE of
$\bmu$ with $\bomega$ takes the form
\begin{displaymath}
  \frac{{\cal D}}{{\cal C}_0} \bmu(x)\bomega = 
  - (8\ln2+\ln|x|^2) \bmu
  + 
  4 \Theta^{\alpha\bar\alpha} d_{\alpha\beta} d_{\bar\alpha\bar\beta} 
  \bnu^{\beta\bar\beta} |x| 
  +
  \cdots\,. 
\end{displaymath}
Other OPEs can be determined similarly and can be found in appendix
\ref{sec:appope}.  
\medskip

Next, we consider the four-point functions of two irreducible and two
reducible representations. The simplest case is the amplitude 
$\langle\bmu\bmu\bomega\bomega\rangle$ which, after imposing the
locality constraints, takes the form
\begin{eqnarray*}
  |z_{12}|^{-\frac12} 
  \langle\bmu\bmu\bomega\bomega\rangle &=& 
  {\cal A}_2 + 
  {\cal A}_1 \ln\left|\frac{z_{13}z_{23}}{z_{12}}\right|^2 + 
  {\cal A}_1 \ln\left|\frac{z_{14}z_{24}}{z_{12}}\right|^2 
  \\*&& + 
  {\cal A}_0 \ln\left|\frac{z_{13}z_{23}}{z_{12}}\right|^2
  \ln\left|\frac{z_{14}z_{24}}{z_{12}}\right|^2
  - 
  {\cal A}_0 \left( 
    \ln\left|
      \frac{\sqrt{z_{13}z_{24}}-\sqrt{z_{14}z_{23}}}
      {\sqrt{z_{13}z_{24}}+\sqrt{z_{14}z_{23}}}
    \right|^2 
  \right)^2 \,,
\end{eqnarray*}
where ${\cal A}_i$ are at this stage arbitrary constants. In order to
determine these we impose the duality relations, 
{\it i.e.} we use the different OPEs to relate this function to
three-point functions which we have already determined. For
example, using the above OPE of $\bmu$ with $\bomega$ 
we find to lowest order in $x$
\begin{eqnarray*}
  \langle\bmu(z_1)\bmu(z_2+x)\bomega(z_2)\bomega(z_3)\rangle &\sim& 
  - (8\ln2+\ln|x|^2) \frac{{\cal C}_0}{{\cal D}}
  \langle\bmu(z_1)\bmu(z_2)\bomega(z_3)\rangle 
  \\*&&
  + 
  4 \frac{{\cal C}_0}{{\cal D}} 
  \Theta^{\alpha\bar\alpha} d_{\alpha\beta} d_{\bar\alpha\bar\beta} 
  \langle\bmu(z_1)\bnu^{\beta\bar\beta}(z_2)\bomega(z_3)\rangle 
  |x| \,.
\end{eqnarray*}
This is to be compared with the same limit of the above four-point
function 
\begin{eqnarray*}
  |z_{12}|^{-\frac12}
  \langle\bmu(z_1)\bmu(z_2+x)\bomega(z_2)\bomega(z_3)\rangle &\sim& 
  {\cal A}_2 + 
  {\cal A}_1 \left(
    \ln|x|^2 + \ln\left|\frac{z_{13}z_{23}}{z_{12}}\right|^2 
  \right) 
  \\*&& + 
  {\cal A}_0 \left(
    \ln|x|^2 \ln\left|\frac{z_{13}z_{23}}{z_{12}}\right|^2 -
    8 \left|\frac{z_{13}}{z_{12}z_{23}}\right| |x|
  \right)\,,
\end{eqnarray*}
and we thus find that 
\begin{displaymath}
  {\cal A}_i = (8\ln2)^i \frac{{\cal C}_0^2}{{\cal D}}
  \qquad\textrm{and}\qquad
  \Theta^{\alpha\bar\alpha} d_{\alpha\beta} d_{\bar\alpha\bar\beta} 
  \Theta^{\beta\bar\beta} = 2\,.
\end{displaymath}
Since $\Theta^{\alpha\bar\alpha} d_{\alpha\beta}
d_{\bar\alpha\bar\beta}  \Theta^{\beta\bar\beta} = 2\det(\Theta)$ the 
second constraint requires $\det(\Theta)=1$ which is what we required
previously (\ref{eq:l10-cond}). 

Similarly, we can determine the other three- and four-point functions
of the fundamental fields, and check their consistency with duality;
some of these amplitudes are explicitly given in appendix 
\ref{sec:appamp}. Unfortunately, the expressions become rather
complicated, and it is not feasible to determine all of the amplitudes
involving arbitrary combinations of the fundamental fields
explicitly. However, we have determined all the relevant three-point
functions (see the above formulae and appendix \ref{sec:appamp}), and
this specifies the theory uniquely. The question reduces then to
whether the theory so specified is indeed consistent; this will be
answered in the affirmative by relating the theory (with this choice
of normalisation constants) to a free field theory that is consistent
\cite{Kausch98}. This free field theory will be described in the
following section.

\section{Symplectic fermions}
\label{sec:fer}

In this section we shall show that the conformal field theory that we
have discussed so far is the bosonic sector of a model of free
``symplectic'' fermions. Here we shall only summarise the essential
features of this fermion model; further details may be found in
\cite{Kausch98} and \cite{Kau95}. 

The chiral algebra of the symplectic fermion model is generated by a
two-component fermion field $\chi^\alpha$ of conformal weight
one whose anti-commutator is given by
\begin{displaymath}
  \{ \chi^\alpha_m, \chi^\beta_n \} = m d^{\alpha\beta}
  \delta_{m+n}\,, 
\end{displaymath}
where $d^{\alpha\beta}$ is the same anti-symmetric tensor as before. 
This algebra has a unique irreducible highest weight representation
generated from a highest weight state $\Omega$ satisfying
$\chi^\alpha_m \Omega = 0$ for $m\geq0$.  We may call this module the
vacuum representation. It contains the triplet $W$-algebra since
\begin{eqnarray*}
  L_{-2} \Omega &=& 
  \frac12 d_{\alpha\beta} \chi^\alpha_{-1} \chi^\beta_{-1} \Omega \,, 
  \\
  W^a_{-3} \Omega &=& 
  t^a_{\alpha\beta} \chi^\alpha_{-2} \chi^\beta_{-1} \Omega 
\end{eqnarray*}
satisfy the triplet algebra \cite{Kausch91,Kau95}. 

Let us consider the maximal generalised highest weight representation
of this chiral algebra that contains the vacuum representation. It 
is freely generated by the negative modes $\chi^\alpha_m, m<0$ from a
four dimensional space of ground states. This space is spanned by two
bosonic states $\Omega$ and $\omega$, and two fermionic states,
$\theta^\alpha$, and the action of the zero-modes $\chi^\alpha_0$ is
given as  
\begin{eqnarray*}
  \chi^\alpha_0 \omega &=& -\theta^\alpha\,, 
  \\
  \chi^\alpha_0 \theta^\beta &=& d^{\alpha\beta} \Omega\,, 
  \\
  L_0 \omega &=& \Omega\,. 
\end{eqnarray*}
Imposing the locality constraints as in section \ref{sec:corr}, the
corresponding non-chiral representation is freely generated by the
negative modes $\chi^\alpha_m, \bar\chi^{\bar\alpha}_m, m<0$ from the
ground space representation 
\begin{displaymath}
  \begin{array}{rcl@{\qquad}rcl}
    \chi^\alpha_0 \bomega &=& -\btheta^\alpha\,, &
    \bar\chi^{\bar\alpha}_0 \bomega &=& -\bar\btheta^{\bar\alpha}\,, 
    \\
    \chi^\alpha_0 \btheta^\beta &=& d^{\alpha\beta} \bOmega\,, &
    \bar\chi^{\bar\alpha}_0 \bar\btheta^{\bar\beta} &=& 
    d^{\bar\alpha\bar\beta} \bOmega\,, 
    \\
    \chi^\alpha_0 \bar\btheta^{\bar\alpha} &=&
    \Theta^{\alpha\bar\alpha} \bOmega\,, &
    \bar\chi^{\bar\alpha}_0 \btheta^\alpha &=&
    -\Theta^{\alpha\bar\alpha} \bOmega\,.
  \end{array}
\end{displaymath}
The space of ground states contains two bosonic states, $\bOmega$ and
$\bomega$, and two fermionic states since the four fermionic states 
$\btheta^\alpha$ and $\bar\btheta^{\bar\alpha}$ are related as 
\begin{displaymath}
  \btheta^\alpha = 
  \Theta^{\alpha\bar\alpha} d_{\bar\alpha\bar\beta}
  \bar\btheta^{\bar\beta}\,, \qquad
  \bar\btheta^{\bar\alpha} = 
  - \Theta^{\alpha\bar\alpha} d_{\alpha\beta} \btheta^\beta\,. 
\end{displaymath}
One can show (see \cite{Kausch98} for further details) that the 
bosonic sector of this representation is isomorphic to the
representation ${\cal R}$. For example, the higher level states of
${\cal R}$ can be expressed as fermionic descendents as 
\begin{eqnarray*}
  \brho^{\alpha\bar\alpha} &=& \chi^\alpha_{-1}
  \bar\btheta^{\bar\alpha}\,, \\
  \bar\brho^{\alpha\bar\alpha} &=& -\bar\chi^{\bar\alpha}_{-1}
  \btheta^\alpha\,, \\
  \bpsi^{\alpha\bar\alpha} &=& 
  \chi^\alpha_{-1}\bar\chi^{\bar\alpha}_{-1}\bOmega\,, \\
  \bphi^{\alpha\bar\alpha} &=& 
  \chi^\alpha_{-1}\bar\chi^{\bar\alpha}_{-1}\bomega\,.
\end{eqnarray*}

The other two representations of the triplet model, the irreducible
representations ${\cal  V}_{-1/8,-1/8}$ and ${\cal V}_{3/8,3/8}$, also
have an interpretation in terms of the symplectic fermion theory: they
correspond to the bosonic sector of the (unique)
$\mathbb{Z}_2$-twisted representation. In this sector, the fermions
are half-integrally moded, but all bosonic operators (including the
triplet algebra generators that are bilinear in the fermions) are
still integrally moded. The ground state of the twisted sector,
$\bmu$, has conformal weight $h=\bar{h}=-1/8$ and satisfies
\begin{displaymath}
  \chi^\alpha_{r}\bmu = \bar\chi^{\bar\alpha}_{r}\bmu = 0\,, 
  \qquad\textrm{for $r>0$}\,, 
\end{displaymath}
while
\begin{displaymath}
  \bnu^{\alpha\bar\alpha} = 
  \chi^\alpha_{-\frac12}\bar\chi^{\bar\alpha}_{-\frac12}\bmu\,. 
\end{displaymath}
With respect to the symplectic fermions, $\bomega$ and $\bmu$ are
cyclic states, and all amplitudes can be reduced to those involving
$\bmu$ and the four ground states of the vacuum representation. This
can be done using the (fermionic) comultiplication formula and its
twisted analogue \cite{Gab97}. The amplitudes involving the ground
states can then be determined by solving differential equations.   
We have determined some of these amplitudes, and we have checked that
they reproduce all the three-point functions of the fundamental fields
of our logarithmic theory.\footnote{We have also checked that some of
the four-point amplitudes agree, for example 
$\langle\bmu\bmu\bomega\bomega\rangle$,
$\langle\bmu\bnu\bomega\bomega\rangle$, 
$\langle\bnu\bnu\bomega\bomega\rangle$ and
$\langle\bomega\bomega\bomega\bomega\rangle$.} This implies that all
amplitudes of the two theories agree, and since the fermion theory is
consistent, thus establishes the consistency of our
logarithmic theory. We should stress that the agreement between the
two set of amplitudes only holds if we make the specific choices for
the normalisation constants (\ref{eq:l00-cond}) and
(\ref{eq:l10-cond}).

\section{Discussion and Conclusion}
\label{sec:res}

Let us first make three comments about the theory we have just
constructed. 

\subsection{Normalisation constants}

In order to give a unified treatment, let us classify the amplitudes
according to their grade, where each irreducible representations
contributes one, and each reducible representation contributes two to
the overall grade. For example, the four-point functions of four
irreducible representations are at grade four, and so are all the two-
and three-point functions that can be derived from them by the use of
the OPE. The grade of the amplitude 
$\langle \bmu \bmu \bomega \bomega\rangle$ is then for example
six, {\it etc}.  

As we have seen in section \ref{sec:amp}, all amplitudes of grade two
are proportional to ${\cal D}$ while all amplitudes of grade four are
proportional to ${\cal C}_0$. It is therefore natural to introduce the
parameters $\Lambda$ and ${\cal O}$ by
\begin{displaymath}
  {\cal D} = \Lambda^2 {\cal O}\,, \qquad
  {\cal C}_0 = \Lambda^4 {\cal O}\,,
\end{displaymath}
so that every amplitude at grade $g$ is proportional to $\Lambda^g
{\cal O}$. Here, $\Lambda$ corresponds to the freedom to rescale the
field $\bmu$, and since we have fixed the normalisations of
$\bnu^{\alpha\bar\alpha},\bomega$ and $\bphi^{\alpha\bar\alpha}$
relative to that of $\bmu$ (see (\ref{mumuope} -- \ref{nunuope})), to
the freedom to rescale all fields with the appropriate power; this
leads to the term $\Lambda^g$ in the normalisation of the amplitude.
The second parameter, ${\cal O}$, can be identified with the
normalisation of the amplitude functional.  In ordinary local
conformal field theories, the parameter ${\cal O}$ is fixed by the
condition that the amplitudes satisfy the {\it cluster property}, {\it
  i.e.} that to leading order every $n+m$-point amplitude is the
product of an $n$- and an $m$-point amplitude. This condition is
essentially equivalent to the uniqueness of the vacuum, {\it i.e.} to
the property that the theory has only one state with vanishing
conformal weight. In our case this condition is not satisfied as there
are two states of conformal weight $0$ in the theory, $\bomega$ and
$\bOmega$. As a consequence, the cluster property does not hold for
any choice of ${\cal O}$, and hence we cannot fix ${\cal O}$ in this
way.

Finally, the last (free) parameter corresponds to the matrix
$\Theta^{\alpha\bar\alpha}$ which describes the coupling between the
left and right $SU(2)$. By performing a global chiral (or anti-chiral)
$SU(2)$ transformation we can always bring this matrix into standard
form, 
\begin{displaymath}
\nonumber
  \Theta^{\alpha\bar\alpha} = d^{\alpha\bar\alpha}\,. 
\end{displaymath}

\subsection{Scale invariance}

It is clear that the logarithms introduce a length scale into
the description and that therefore manifest scale invariance is
broken. However, the amplitudes are in fact invariant under the
dilatations $z_i\mapsto\e^\lambda z_i$ if the corresponding
transformation on the states (apart from the usual factor of
$\exp(\lambda(h+\bar h))$) is given as 
\begin{displaymath}
  \bomega \mapsto \bomega + 2\lambda \bOmega \,, 
  \qquad
  \bphi^{\alpha\bar\alpha} \mapsto 
  \bphi^{\alpha\bar\alpha} + 2\lambda \bpsi^{\alpha\bar\alpha}\,. 
\end{displaymath}
This is a direct consequence of the fact that $L_0$ can be identified
with the scale operator, and that $L_0$ does not act diagonally on
$\bomega$ and $\bphi$. It is also easy to check the scale invariance
of the amplitudes explicitly.

\subsection{Partition function}

We can also read off immediately the partition function of the
resulting theory. First of all, the characters of the non-chiral
irreducible representations are simply the product of a left and right
chiral character
\begin{eqnarray*}
  \chi_{{\cal V}_{-1/8,-1/8}}(\tau) &=& 
  \chi_{{\cal V}_{-1/8}}(\tau) \bar\chi_{{\cal V}_{-1/8}}(\bar\tau) = 
  \left| \eta(\tau)^{-1} \theta_{0,2}(\tau) \right|^2 \,, 
  \\*
  \chi_{{\cal V}_{3/8,3/8}}(\tau) &=& 
  \chi_{{\cal V}_{3/8}}(\tau) \bar\chi_{{\cal V}_{3/8}}(\bar\tau) = 
  \left| \eta(\tau)^{-1} \theta_{2,2}(\tau) \right|^2\,. 
\end{eqnarray*}
To determine the character of the reducible representation ${\cal R}$,
let us recall that each vertex in the diagrammatical representation of
${\cal R}$ corresponds to an irreducible representation of the left
and right triplet algebra. Putting the different contributions
together we find
\begin{eqnarray*}
  \chi_{{\cal R}}(\tau) &=& 
  2 \chi_{{\cal V}_0}(\tau) \bar\chi_{{\cal V}_0}(\bar\tau) + 
  2 \chi_{{\cal V}_1}(\tau) \bar\chi_{{\cal V}_0}(\bar\tau) + 
  2 \chi_{{\cal V}_0}(\tau) \bar\chi_{{\cal V}_1}(\bar\tau) + 
  2 \chi_{{\cal V}_1}(\tau) \bar\chi_{{\cal V}_1}(\bar\tau)
  \\*
  &=&
  2 \left| \chi_{{\cal V}_0}(\tau) + \chi_{{\cal V}_1}(\tau) \right|^2
  \\*
  &=&
  2 \left| \eta(\tau)^{-1} \theta_{1,2}(\tau) \right|^2 \,,
\end{eqnarray*}
where ${\cal V}_0$ and ${\cal V}_1$ denote the irreducible
representations with conformal weights $0$ and $1$, respectively
\cite{GKau96b}. The partition function of the full theory is thus 
\begin{displaymath}
  Z = 
  \chi_{{\cal V}_{-1/8,-1/8}}(\tau) + 
  \chi_{{\cal V}_{3/8,3/8}}(\tau) + 
  \chi_{{\cal R}}(\tau)
  =
  \left|\eta(\tau)\right|^{-2} 
  \sum_{k=0}^3 \left|\theta_{k,2}(\tau)\right|^2 \,,
\end{displaymath}
and this is indeed modular invariant. Actually, it is the same
partition function as that of a free boson compactified on a circle of
radius $\sqrt{2}$ \cite{Ginsparg}. However, in our case the partition
function is not simply the sum of products of left- and right- chiral
representations of the chiral algebra as the non-chiral
representation ${\cal R}$ is {\it not} the tensor product of a left-
and right-chiral representation, but only a quotient thereof. As we
have explained before, this follows directly from locality.

\subsection{Conclusions}

We have constructed a consistent local logarithmic
theory, the first such theory to be understood in detail. 
Schematically speaking, its (non-chiral) fusion rules are described by
\begin{eqnarray*}
\bmu \otimes \bmu &= & \bomega \\
\bmu \otimes \bnu & = & \bomega \\
\bnu \otimes \bnu & = & \bomega \\ 
\bmu \otimes \bomega & = & \bmu \oplus \bnu \\
\bnu \otimes \bomega & = & \bmu \oplus \bnu \\
\bomega \otimes \bomega & = & 2 \bomega \,.
\end{eqnarray*}
We have constructed the two-, three- and some of the four-point
amplitudes of the fundamental fields, and have shown that they satisfy
the locality and duality constraints. We have also shown
that this theory is equivalent to a the bosonic subtheory of a free
fermionic theory, thereby establishing that it is indeed
consistent. We have determined the partition function of the theory,
and shown that it is indeed modular invariant. Apart from the
appearance of logarithms in some of the correlation functions, this
theory defines a bona fide local conformal field theory.

\paragraph{Acknowledgements:}

We would like to thank Wolfgang Eholzer, Michael Flohr, Peter Goddard,
G\'erard Watts for useful discussions. 

M.R.G. is grateful to Jesus College, Cambridge, for a Research
Fellowship. This work has also been supported in part by PPARC.

\section*{Appendix}

\appendix

\section{Operator product expansions}
\label{sec:appope}

This appendix lists the relevant operator product expansions.  Terms
up to order $O(x,\bar x)$ for states in $\mathcal{R}$ and up to order
$O(x^{1/2}, \bar x^{1/2})$ for states in $\mathcal{V}_{-1/8,-1/8}$ and
$\mathcal{V}_{3/8,3/8}$ are given. We have included the arbitrary
constant $\Lambda$ whose significance is explained in section
\ref{sec:res}. 
\begin{eqnarray*}
  %
  |x|^{-\frac12} \bmu(x)\bmu &=& 
  \left(
    \bomega + \ln|x|^2 \bOmega 
  \right)
  +
  \frac12 L_{-1} \bomega x
  +
  \frac12 \bar L_{-1} \bomega \bar x
  +
  \frac14 L_{-1} \bar L_{-1} \bomega |x|^2
  +
  \cdots,
  \\
  |x|^{\frac12} \bmu(x)\bnu^{\alpha\bar\alpha} &=& 
  - \Theta^{\alpha\bar\alpha} \bOmega
  + \frac12 \brho^{\alpha\bar\alpha} x
  + \frac12 \bar\brho^{\alpha\bar\alpha} \bar{x}
  \\*&&{}
  + \frac14 \left(
    \bphi^{\alpha\bar\alpha} 
    +
    (\ln|x|^2+2) \bpsi^{\alpha\bar\alpha} 
  \right) |x|^2
  +
  \cdots
  \\
  |x|^{\frac32} \bnu^{\alpha\bar\alpha}(x)\bnu^{\beta\bar\beta} &=& 
  -\frac14 d^{\alpha\beta} d^{\bar\alpha\bar\beta}
  \left(\bomega + (\ln|x|^2+4) \bOmega\right)
  \\*&&{}
  +
  \left(
    \frac18 d^{\alpha\beta} d^{\bar\alpha\bar\beta} L_{-1} \bomega
    +
    \frac14 \Theta^{\beta\bar\eta} d^{\bar\alpha\bar\beta}
    d_{\bar\eta\bar\gamma} \brho^{\alpha\bar\gamma} 
  \right) x
  \\*&&{}
  +
  \left(
    \frac18 d^{\alpha\beta} d^{\bar\alpha\bar\beta} \bar L_{-1} \bomega
    +
    \frac14 \Theta^{\eta\bar\beta} d^{\alpha\beta}
    d_{\eta\gamma} \brho^{\gamma\bar\alpha} 
  \right) \bar x
  \\*&&{}
  -
  \left(
    \frac{3}{16} d^{\alpha\beta} d^{\bar\alpha\bar\beta} 
    \tilde\Theta_{\bar\gamma\gamma} \bpsi^{\gamma\bar\gamma}
    +
    \frac14 \Theta^{\alpha\bar\alpha} \bpsi^{\beta\bar\beta}
    +
    \frac34 \Theta^{\beta\bar\beta} \bpsi^{\alpha\bar\alpha}
  \right) |x|^2
  +
  \cdots, 
  \\[\bigskipamount]
  %
  %
  \Lambda^{-2} \bmu(x)\bomega &=& 
  - (8\ln2+\ln|x|^2) \bmu
  + 
  4 \tilde\Theta_{\bar\alpha\alpha} \bnu^{\alpha\bar\alpha} |x| 
  +
  \cdots, 
  \\
  \Lambda^{-2} |x| \bnu(x)^{\alpha\bar\alpha}\bomega &=& 
  - \Theta^{\alpha\bar\alpha} \bmu 
  - (8\ln2 - 4 + \ln|x|^2) \bnu^{\alpha\bar\alpha} |x| 
  +
  \cdots, 
  \\
  \Lambda^{-2} |x|^2 \bmu(x)\bphi^{\alpha\bar\alpha} &=& 
  -\frac14 \Theta^{\alpha\bar\alpha} \bmu
  - (8\ln2 - 2 + \ln|x|^2) \bnu^{\alpha\bar\alpha} |x|
  +
  \cdots, 
  \\
  \Lambda^{-2} |x|^3 \bnu^{\alpha\bar\alpha}(x)\bphi^{\beta\bar\beta} &=& 
  \frac14 d^{\alpha\beta} d^{\bar\alpha\bar\beta} 
  (8\ln2 + 2 + \ln|x|^2) \bmu
  \\*&&
  - 
  \left(
    3 d^{\beta\gamma} d^{\bar\beta\bar\gamma}
    \tilde\Theta_{\bar\eta\eta} 
    - 
    \frac34 \Theta^{\gamma\bar\gamma} \delta^\beta_\eta
  \delta^{\bar\beta}_{\bar\eta} 
  \right) 
  \bnu^{\eta\bar\eta} |x|
  +
  \cdots, 
  \\[\bigskipamount]
  %
  %
  \Lambda^{-2} \bomega(x) \bomega &=&
  - \left( 4\ln2+\ln|x|^2\right)
  \Biggl[
    2 \bomega + 
    ( 4\ln2+\ln|x|^2) \bOmega
    \\*&&\qquad{}
    +
    \tilde\Theta_{\bar\alpha\alpha} \brho^{\alpha\bar\alpha} x
    +
    \tilde\Theta_{\bar\alpha\alpha} \bar\brho^{\alpha\bar\alpha}
    \bar x
  \Biggr]
  +
  \tilde\Theta_{\bar\alpha\alpha} \bphi^{\alpha\bar\alpha} |x|^2
  +
  \cdots, 
  \\
  \Lambda^{-2} |x|^2 \bphi(x)^{\alpha\bar\alpha}\bomega &=& 
  - \Theta^{\alpha\bar\alpha} \bomega
  \\*&&{}
  - \left(
    \Theta^{\alpha\bar\alpha} L_{-1}\bomega
    - 
    ( 4\ln2 + \ln|x|^2) \brho^{\alpha\bar\alpha} 
  \right) x
  \\*&&{}
  - \left(
    \Theta^{\alpha\bar\alpha} \bar L_{-1}\bomega
    - 
    ( 4\ln2 + \ln|x|^2) \bar\brho^{\alpha\bar\alpha} 
  \right) \bar x
  \\*&&{}
  - \Bigl(
    (4\ln2 + \ln|x|^2) (4\ln2 - 2 + \ln|x|^2) \bpsi^{\alpha\bar\alpha} 
     \\*&&\quad{}
     +
     2 (4\ln2 - 1 + \ln|x|^2) \bphi^{\alpha\bar\alpha} 
     +
    \Theta^{\alpha\bar\alpha} \tilde\Theta_{\bar\eta\eta} 
    \bpsi^{\eta\bar\eta} 
  \Bigr) |x|^2 +  \cdots\,, 
  \\
  \Lambda^{-2} |x|^4 \bphi(x)^{\alpha\bar\alpha}\bphi^{\beta\bar\beta} &=& 
  \Theta^{\alpha\bar\alpha} \Theta^{\beta\bar\beta} \bOmega 
  \\*&&{}
  +
  d^{\alpha\beta} d^{\bar\alpha\bar\beta} \Biggl[
    2 ( 4\ln2 + 1 + \ln|x|^2) \bomega
    \\*&&\qquad{}
    +
    ( 4\ln2 + \ln|x|^2) ( 4\ln2 + 2 + \ln|x|^2) \bOmega
  \Biggr]
  \\*&&{}
  +
  2 \left(
    (4\ln2 + \ln|x|^2)
    \Theta^{\gamma\bar\beta} d^{\alpha\beta}
    d_{\gamma\eta} \bar\brho^{\eta\bar\alpha}
    +
    \Theta^{\beta\bar\gamma} d^{\bar\alpha\bar\beta}
    d_{\bar\gamma\bar\eta} \bar\brho^{\alpha\bar\eta}
  \right) \bar x
  \\*&&{}
  +
  2 \left(
    (4\ln2 + \ln|x|^2)
    \Theta^{\gamma\bar\beta} d^{\alpha\beta}
    d_{\gamma\eta} \bar\brho^{\eta\bar\alpha}
    +
    \Theta^{\beta\bar\gamma} d^{\bar\alpha\bar\beta}
    d_{\bar\gamma\bar\eta} \bar\brho^{\alpha\bar\eta}
  \right) \bar x
  \\*&&{}
  -
  2 
  \Biggl[
    d^{\alpha\beta} d^{\bar\alpha\bar\beta}
    \tilde\Theta_{\bar\eta\eta} 
    \left(
      2 \bphi^{\eta\bar\eta} + (4\ln2+\ln|x|^2) \bpsi^{\eta\bar\eta} 
    \right) 
    \\*&&\qquad{}
    +
    \Theta^{\alpha\bar\alpha} \bpsi^{\beta\bar\beta}
    -
    (4\ln2+\ln|x|^2) \Theta^{\beta\bar\beta}
    \bpsi^{\alpha\bar\alpha} 
  \biggr] |x|^2
  +
  \cdots. 
\end{eqnarray*}
The operator product of $\bOmega$ with any field ${\bf S}$ is simply
given by $\bOmega(x) {\bf S} = \Lambda^2 {\bf S}$ to all orders. For
$\Lambda^2=1$, $\bOmega$ can be thought of as the unit operator,
except that its one-point function vanishes, 
$\langle \bOmega \rangle =0$.

\section{Amplitudes}
\label{sec:appamp}

In this appendix we list some of the higher grade amplitudes of the
theory that we have checked to be consistent with the OPEs and the
lower grade amplitudes determined before. For simplicity we have set
$\Lambda={\cal O}=1$. These parameters can be restored by multiplying
an amplitude of grade $g$ by a factor of $\Lambda^g {\cal O}$.

The fundamental 3-point amplitudes of reducible representations are
\begin{eqnarray*}
  \langle\bomega\bomega\bomega\rangle &=& 
  48 (\ln2)^2
  + 
  8\ln2 \ln|z_{12}z_{13}z_{23}|^2
  \\*&&+
  2 \left(
    \ln|z_{12}|^2 \ln|z_{13}|^2 + 
    \ln|z_{12}|^2 \ln|z_{23}|^2 + 
    \ln|z_{13}|^2 \ln|z_{23}|^2
  \right)
  \\*&&-
  \left(
    (\ln|z_{12}|^2)^2 + 
    (\ln|z_{13}|^2)^2 + 
    (\ln|z_{23}|^2)^2
  \right) \,,
  \\
  \langle\bphi^{\alpha\bar\alpha}\bomega\bomega\rangle &=& 
  \Theta^{\alpha\bar\alpha}   \Biggl[
  \frac{1}{|z_{12}|^2}
  \left(
    4\ln2 + \ln\left|\frac{z_{13}z_{23}}{z_{12}}\right|^2
  \right)
  \\*&&\quad{}
  +
  \frac{1}{|z_{13}|^2}
  \left(
    4\ln2 + \ln\left|\frac{z_{12}z_{23}}{z_{13}}\right|^2
  \right)
  \\*&&\quad{}
  +
  \left|\frac{z_{23}}{z_{12}z_{13}}\right|^2 
  \left(
    4\ln2 + \ln\left|\frac{z_{12}z_{13}}{z_{23}}\right|^2
  \right)
  \Biggr]\,, 
  \\
  \langle\bphi^{\alpha\bar\alpha}\bphi^{\beta\bar\beta}\bomega\rangle &=&
  \Theta^{\alpha\bar\alpha} \Theta^{\beta\bar\beta}   
  \left|
    \frac{1}{z_{12}^2} + \frac{1}{z_{13}z_{23}}
  \right|^2
  \\*&& \quad +
  d^{\alpha\beta} d^{\bar\alpha\bar\beta}  
  \frac{1}{|z_{12}|^4}
  \Biggl[
    \left(
      4\ln2 - 2 + \ln\left|\frac{z_{13}z_{23}}{z_{12}}\right|^2
    \right)
    \\*&&\qquad{}\times
    \left(
      4\ln2 + \ln\left|\frac{z_{13}z_{23}}{z_{12}}\right|^2 +
      \frac{z_{12}^2}{z_{13}z_{23}} + 
      \frac{\bar z_{12}^2}{\bar z_{13}\bar z_{23}}
    \right)
    \\*&&\quad{}
    -
    \left(
      4\ln2 + 2\ln\left|z_{13}\right|^2 +
      \frac{z_{12}}{z_{23}} + \frac{\bar z_{12}}{\bar z_{23}}
    \right)
    \\*&&\qquad{}\times
    \left(
      4\ln2 + 2\ln\left|z_{23}\right|^2 -
      \frac{z_{12}}{z_{13}} - \frac{\bar z_{12}}{\bar z_{13}}
    \right)
  \Biggr]\,, 
  \\
  \langle\bphi^{\alpha\bar\alpha}\bphi^{\beta\bar\beta}
  \bphi^{\gamma\bar\gamma}\rangle &=&
  - \frac{{\cal T}_1}{|z_{12}z_{13}z_{23}|^2}
  \\*&&\quad\times
  \Biggr[
    \left|\frac{z_{13}+z_{23}}{z_{13}z_{23}}\right|^2 
    \left(
      4\ln2 + 1 +\ln\left|\frac{z_{13}z_{23}}{z_{12}}\right|^2
    \right)
    \\*&&\qquad
    +
    \left|\frac{z_{12}-z_{23}}{z_{12}z_{23}}\right|^2 
    \left(
      4\ln2 + 1 +\ln\left|\frac{z_{12}z_{23}}{z_{13}}\right|^2
    \right)
    \\*&&\qquad
    +
    \left|\frac{z_{12}+z_{13}}{z_{12}z_{13}}\right|^2 
    \left(
      4\ln2 + 1 +\ln\left|\frac{z_{12}z_{13}}{z_{23}}\right|^2
    \right)
  \Biggr]
  \\*&&
  - \frac12 
  \frac{(|z_{12}|^2+|z_{13}|^2+|z_{23}|^2)
    (z_{13}\bar z_{23}-z_{23}\bar z_{13})}
  {|z_{12}z_{13}z_{23}|^4}
  \\*&&\quad\times
  \left[
    {\cal T}_2 (z_{13}\bar z_{23}-z_{23}\bar z_{13})
    +
    {\cal T}_3 (|z_{12}|^2+|z_{13}|^2+|z_{23}|^2)
  \right]
  \,,
\end{eqnarray*}
where
\begin{eqnarray*}
  {\cal T}_1 &=& 
  \Theta^{\alpha\bar\alpha} d^{\beta\gamma} d^{\bar\beta\bar\gamma} 
  z_{13} \bar z_{13} 
  -
  \Theta^{\alpha\bar\beta} d^{\beta\gamma} d^{\bar\alpha\bar\gamma} 
  z_{13} \bar z_{23} 
  -
  \Theta^{\beta\bar\alpha} d^{\alpha\gamma} d^{\bar\beta\bar\gamma} 
  z_{23} \bar z_{13} 
  +
  \Theta^{\beta\bar\beta} d^{\alpha\gamma} d^{\bar\alpha\bar\gamma} 
  z_{23} \bar z_{23} 
  \,,
  \\
  {\cal T}_2 &=& 
  \Theta^{\alpha\bar\alpha} d^{\beta\gamma}
  d^{\bar\beta\bar\gamma} 
  +
  \Theta^{\beta\bar\beta} d^{\alpha\gamma}
  d^{\bar\alpha\bar\gamma} 
  +
  \Theta^{\gamma\bar\gamma} d^{\alpha\beta}
  d^{\bar\alpha\bar\beta} 
  \,,
  \\
  {\cal T}_3 &=& 
  \Theta^{\alpha\bar\beta} d^{\beta\gamma}
  d^{\bar\alpha\bar\gamma} 
  -
  \Theta^{\beta\bar\alpha} d^{\alpha\gamma}
  d^{\bar\beta\bar\gamma} 
  \,.
\end{eqnarray*}
Using the tensor identities in appendix \ref{sec:appten}, it is not
difficult to see that ${\cal T}_1$ and ${\cal T}_2$ are completely
symmetric under the exchange of any two of the three fields, whereas 
${\cal T}_3$ is completely anti-symmetric. Furthermore, it is easy to
check that $(z_{13} \bar{z}_{23} - z_{23} \bar{z}_{13})$ is completely
anti-symmetric, and it thus follows that the above amplitude is
completely symmetric. 
\bigskip

The four-point amplitudes of grade four were already given for a
certain domain in the cross-ratio $x$ in section \ref{sec:amp}. Other
regions of $x$ are described by different orderings of the fields,
where the corresponding cross-ratio is again small
\begin{eqnarray*}
  \langle\bmu(z_1)\bnu^{\alpha\bar\alpha}(z_2)\bmu(z_3)
  \bnu^{\beta\bar\beta}(z_4)\rangle 
  &=& 
  -\frac{\pi}{4} d^{\alpha\beta} d^{\bar\alpha\bar\beta}
  \left|
    \frac{z_{13}^3}{z_{12}z_{14}z_{24}z_{23}z_{34}}
  \right|^{\frac12} \\
  & & \qquad\qquad \times 
  \left[
    D_1(x)\tilde D_1(\bar x) + \tilde D_1(x)D_1(\bar x)
  \right] \,,
  \\
  \langle\bmu(z_1)\bnu^{\alpha\bar\alpha}(z_2)
  \bnu^{\beta\bar\beta}(z_3)\bmu(z_4)\rangle 
  &=& 
  \frac{\pi}{4} d^{\alpha\beta} d^{\bar\alpha\bar\beta}
  \left|
    \frac{z_{13}z_{14}z_{24}}{z_{12}z_{23}^3z_{34}}
  \right|^{\frac12}
  \left[
    D_2(x)\tilde E(\bar x) +  \tilde E(x)D_2(\bar x)
  \right]
  \,.
\end{eqnarray*}
It is easy to check (remembering that we have set ${\cal C}_1=1$) that
the above four-point amplitudes give rise to the same three-point
amplitudes as the ones determined in section \ref{sec:amp}.
\medskip

\noindent Some four-point amplitudes of grade six are
\begin{eqnarray*}
  |z_{12}|^{-\frac12} 
  \langle\bmu\bmu\bomega\bomega\rangle &=& 
  \left( 
    8\ln2 + \ln\left|\frac{z_{13}z_{23}}{z_{12}}\right|^2
  \right) \left( 
    8\ln2 + \ln\left|\frac{z_{14}z_{24}}{z_{12}}\right|^2
  \right)
  - H(x,\bar x)^2 \,,
  \\
  |z_{12}|^{\frac12} 
  \langle\bmu\bnu^{\alpha\bar\alpha}\bomega\bomega\rangle &=& 
  \Theta^{\alpha\bar\alpha} \Biggl[
  \left|\frac{z_{13}}{z_{23}}\right| 
  \left(
    8\ln2 + \ln\left|\frac{z_{14}z_{24}}{z_{12}}\right|^2
  \right)
  \\*&&\qquad \quad +
  \left|\frac{z_{14}}{z_{24}}\right| 
  \left(
    8\ln2 + \ln\left|\frac{z_{13}z_{23}}{z_{12}}\right|^2
  \right)
  \\*&&\qquad \quad +
  \left(
    \sqrt{\frac{z_{13}\bar z_{14}}{z_{23}\bar z_{24}}} + 
    \sqrt{\frac{z_{14}\bar z_{13}}{z_{24}\bar z_{23}}} 
  \right)
  H(x,\bar x)
  \Biggr] \,, 
  \\
  |z_{12}|^{\frac32} 
  \langle\bnu^{\alpha\bar\alpha}\bnu^{\beta\bar\beta}\bomega\bomega\rangle &=& 
  \Theta^{\alpha\bar\alpha} \Theta^{\beta\bar\beta} 
  \left|
    \sqrt{\frac{z_{13}z_{24}}{z_{14}z_{23}}} 
    - 
    \sqrt{\frac{z_{14}z_{23}}{z_{13}z_{24}}} 
  \right|^2
  \\*&&\quad{}-
  \frac14 d^{\alpha\beta} d^{\bar\alpha\bar\beta} 
  \Biggl[
  \left( 
    8\ln2 - 4 + \ln\left|\frac{z_{13}z_{23}}{z_{12}}\right|^2
  \right) 
  \\*&&\qquad{}\times
  \left( 
    8\ln2 - 4 + \ln\left|\frac{z_{14}z_{24}}{z_{12}}\right|^2
  \right)
  \\*&&\quad{}
  -
  \left(
    H(x,\bar x) 
    + 
    2 \sqrt{\frac{z_{13}z_{24}}{z_{14}z_{23}}}
    + 
    2 \sqrt{\frac{\bar z_{13}\bar z_{24}}{\bar z_{14}\bar z_{23}}}
  \right)
  \\*&&\qquad{}\times
  \left(
    H(x,\bar x) 
    + 
    2 \sqrt{\frac{z_{14}z_{23}}{z_{13}z_{24}}}
    + 
    2 \sqrt{\frac{\bar z_{14}\bar z_{23}}{\bar z_{13}\bar z_{24}}}
  \right)
  \Biggr]\,, 
\end{eqnarray*}
where 
\begin{displaymath}
  H(x,\bar x) =   
  \ln\left|
    \frac{1-\sqrt{1-x}}{1+\sqrt{1-x}}
  \right|^2
  = 
  \ln\left|
    \frac{\sqrt{z_{13}z_{24}}-\sqrt{z_{14}z_{23}}}
    {\sqrt{z_{13}z_{24}}+\sqrt{z_{14}z_{23}}}
  \right|^2 \,.
\end{displaymath}

The fundamental four-point amplitude of grade eight is 
\begin{eqnarray*}
  \langle\bomega\bomega\bomega\bomega\rangle &=& 
  256(\ln2)^3 + 32(\ln2)^2 ({\circ}{-}{\circ})
  + 
  8\ln2 \Bigl[
    ({\circ}{-}{\circ}{-}{\circ}) - 
    ({\circ}{=}{\circ})
  \Bigr] 
  \\*&&+
  \Bigl[
    2({\circ}{-}{\circ}{-}{\circ}{-}{\circ}) - 
    2({\circ}{=}{\circ}{\circ}{-}{\circ}) - 
    2(\bigtriangledown)
  \Bigr] \,,
\end{eqnarray*}
where
\begin{eqnarray*}
  ({\circ}{-}{\circ}) &=& \sum_{ij} \ln|z_{ij}|^2\,, 
  \\
  ({\circ}{-}{\circ}{-}{\circ}) &=& 
  \sum_{ijk} \ln|z_{ij}|^2 \ln|z_{jk}|^2\,, 
  \\
  ({\circ}{-}{\circ}{-}{\circ}{-}{\circ}) &=& 
  \sum_{ijkl} \ln|z_{ij}|^2 \ln|z_{jk}|^2 \ln|z_{kl}|^2\,, 
  \\
  ({\circ}{=}{\circ}) &=& \sum_{ij} \left(\ln|z_{ij}|^2\right)^2\,, 
  \\
  ({\circ}{=}{\circ}{\circ}{-}{\circ}) &=& 
  \sum_{ijkl} \left(\ln|z_{ij}|^2\right)^2 \ln|z_{kl}|^2\,, 
  \\
  (\bigtriangledown) &=& 
  \sum_{ijk} \ln|z_{ij}|^2 \ln|z_{jk}|^2 \ln|z_{ki}|^2\,. 
\end{eqnarray*}
The sums are over pairwise distinct labelled graphs ({\it i.e.} graphs
with vertices); labelled graphs that differ by a graph symmetry are
only counted once.

\section{Elliptic Integrals}
\label{sec:appell}

The four-point functions can be expressed in terms of complete
elliptic integrals, $K$, $\tilde K$, $E$, $\tilde E$. These are
related to the hypergeometric series near $0$ or $1$,
\begin{displaymath}
  K(x) = {}_2F_1(1/2,1/2;1;x)\,, \qquad
  E(x) = {}_2F_1(-1/2,1/2;1;x)\,.
\end{displaymath}
\begin{displaymath}
  \tilde K(x) = K(1-x)\,, \qquad
  \tilde E(x) = E(1-x)\,.
\end{displaymath}
The other functions appearing in four-point functions are
\begin{eqnarray*}
  D_1(x) &=& E(x) - (1-x) K(x) = 2x(1-x) K'(x)\,, \\
  D_2(x) &=& -E(x) + K(x) = -2x E'(x)\,, \\
  \tilde D_1(x) &=& \tilde E(x) - x \tilde K(x)\,, \\
  \tilde D_2(x) &=& -\tilde E(x) + \tilde K(x)\,, \\
  F_1(x) &=& 2xE(x) - x(2-x)K(x)\,, \\
  F_2(x) &=& (2-x)E(x) - \frac12 (2-2x+x^2)K(x)\,, \\
  \tilde F_1(x) &=& 2x\tilde E(x) - x^2\tilde K(x)\,, \\
  \tilde F_2(x) &=& (2-x)\tilde E(x) - \frac12 (2-x^2)\tilde K(x) \,.
\end{eqnarray*}
The functions with a $(\tilde\cdot)$ have a logarithmic branch cut  
at zero,which can be seen from the analytic continuation for 
$|\arg x|<\pi$ (\textit{c.f.} Erd\'eli \textit{et.al.} \cite{EMOT53}),
\begin{eqnarray*}
  K(1-x) &=& -\frac1\pi \left[ \ln(x/16) K(x) + 2 M(x) \right]\,, \\
  E(1-x) &=& -\frac1\pi \left[ \ln(x/16) D_2(x) - 2 N(x) \right]\,,
\end{eqnarray*}
where
\begin{eqnarray*}
  M(x) &=& \sum_{n=1}^\infty \frac{(1/2)_n^2}{n!^2} h_n x^n\,, \\
  N(x) &=& 1+\sum_{n=1}^\infty \frac{(-1/2)_n(1/2)_n}{n!(n+1)!}
  (h_n+h_{n-1}) x^n\,,
\end{eqnarray*}
and
\begin{displaymath}
  h_n = \psi(1) - \psi(n+1) - \psi(1/2) + \psi(n+1/2)\,.
\end{displaymath}
Furthermore, for analytic continuation to infinity one has
\begin{displaymath}
  \begin{array}{rcl@{\qquad}rcl}
    x^{-1/2} K(1/x) &=& K(x) + i\tilde K(x)\,, &
    x^{-1/2} \tilde K(1/x) &=& \tilde K(x)\,, \\
    x^{1/2} E(1/x) &=& D_1(x) + i\tilde D_1(x)\,, &
    x^{1/2} \tilde E(1/x) &=& \tilde E(x)\,.
  \end{array}
\end{displaymath}

\section{$su(2)$ tensors.}
\label{sec:appten}

We choose a Cartan-Weyl basis for $su(2)$ so that the non-vanishing
structure constants are
\begin{displaymath}
  f^{0\pm}_\pm = \pm1\,, \qquad f^{\pm\mp}_0 = \pm2\,,
\end{displaymath}
and the metric is given by
\begin{displaymath}
  g^{00} = 1\,, \qquad g^{\pm\mp} = 2\,.
\end{displaymath}
The spin $1/2$ representation is given by the matrices
$(t^a)^\alpha_\beta$ whose non-vanishing entries are
\begin{displaymath}
  t^{0\pm}_\pm = \pm\frac12\,, \qquad t^{\pm\mp}_\pm = 1\,.
\end{displaymath}
The anti-symmetric tensor $d^{\alpha\beta}$ and its inverse
$d_{\alpha\beta}$, normalised to $d^{\pm\mp} = d_{\mp\pm} = \pm1$, can
be used to raise and lower indices in the spin $1/2$ representation.
The other tensors appearing in correlation functions are
\begin{eqnarray*}
  t^{a\alpha\beta}: && 
  t^{a\alpha\beta} = t^{a\alpha}_\gamma d^{\gamma\beta}, \\
  &&
  t^{0\pm\mp} = \frac12\,, \qquad t^{\pm\mp\mp} = \pm1\,, \\
  t^a_{\alpha\beta}: &&
  t^a_{\alpha\beta} = t^{a\gamma}_\alpha d_{\gamma\beta}, \\ 
  &&
  t^0_{\pm\mp} = -\frac12\,, \qquad t^\pm_{\pm\pm} = \pm1\,, \\
  h^{\alpha\beta\gamma}_\eta: &&
  h^{\alpha\beta\gamma}_\eta =
  2t^{a\alpha\beta}g_{ab}t^{b\gamma}_\eta = 
  -\frac12 \left(
    d^{\alpha\gamma} \delta^\beta_\eta + 
    d^{\beta\gamma} \delta^\alpha_\eta 
  \right), \\ 
  && 
  h^{\pm\pm\mp}_\pm = \mp1\,, \qquad
  h^{+-\pm}_\pm = h^{-+\pm}_\pm = \pm\frac12\,, \\
  h^{\alpha\beta\gamma\delta}: &&
  h^{\alpha\beta\gamma\delta} =
  2t^{\alpha\alpha\beta}g_{ab}t^{b\gamma\delta} = 
  -\frac12 \left(
    d^{\alpha\gamma} d^{\beta\delta} + 
    d^{\alpha\delta} d^{\beta\gamma} 
  \right), \\ 
  &&
  h^{\pm\pm\mp\mp} = -1\,, \qquad
  h^{\pm\mp\pm\mp} = h^{\pm\mp\mp\pm} = \frac12\,.
\end{eqnarray*}
In addition the following tensor relations hold:
\begin{eqnarray*}
  d^{\alpha\beta} d^{\gamma\delta} - 
  d^{\alpha\gamma} d^{\beta\delta} + 
  d^{\alpha\delta} d^{\beta\gamma} &=& 0, 
  \\
  \Theta^{\alpha\bar\alpha} \Theta^{\beta\bar\beta} - 
  \Theta^{\alpha\bar\beta} \Theta^{\beta\bar\alpha} - 
  d^{\alpha\beta} d^{\bar\alpha\bar\beta} &=& 0, 
  \\
  \Theta^{\alpha\bar\alpha} d^{\beta\gamma} + 
  \Theta^{\beta\bar\alpha} d^{\gamma\alpha} + 
  \Theta^{\gamma\bar\alpha} d^{\alpha\beta} &=& 0, 
  \\
  \Theta^{\alpha\bar\alpha} d^{\bar\beta\bar\gamma} + 
  \Theta^{\alpha\bar\beta} d^{\bar\gamma\bar\alpha} + 
  \Theta^{\alpha\bar\gamma} d^{\bar\alpha\bar\beta} &=& 0. 
\end{eqnarray*}

\end{document}